\documentclass[sigconf]{acmart}
\pdfoutput=1
\copyrightyear{2021} 
\acmYear{2021} 
\setcopyright{acmlicensed}
\acmConference[CCS '21] {Proceedings of the 2021 ACM SIGSAC Conference on Computer and Communications Security}{November 15--19, 2021}{Virtual Event, Republic of Korea.}
\acmBooktitle{Proceedings of the 2021 ACM SIGSAC Conference on Computer and Communications Security (CCS '21), November 15--19, 2021, Virtual Event, Republic of Korea}
\acmPrice{15.00}
\acmISBN{978-1-4503-8454-4/21/11} 
\acmDOI{10.1145/3460120.3485253}

\usepackage{xspace}
\usepackage{color}
\usepackage{url}
\usepackage{subcaption}
\usepackage{paralist}
\usepackage{wrapfig}
\usepackage{multirow}
\usepackage{listings}
\usepackage{booktabs}
\usepackage{makecell}
\usepackage{boxedminipage}

\usepackage{graphicx}
\usepackage{minibox}
\usepackage[ruled]{algorithm2e}
\usepackage{pifont}
\usepackage{color}
\usepackage{xcolor}
\usepackage{subfiles}
\usepackage[utf8]{inputenc}
\usepackage[english]{babel}

\usepackage{grffile}

\usepackage[all]{nowidow}

\newcounter{BalanceAtReference}
\newcounter{ReferenceIndexForBalancing}

\makeatletter

\global\@ACM@balancefalse

\def\@balancelastpageonce{%
  \ifnum\value{ReferenceIndexForBalancing}=\value{BalanceAtReference}
    \newpage
  \else
    \relax
  \fi
  \stepcounter{ReferenceIndexForBalancing}
}
\pretocmd{\bibitem}{\@balancelastpageonce}
  {} 
  {\@latex@error{Patching \bibitem failed}{\@ehd}}

\makeatother

\definecolor{Maroon}{rgb}{0.6, 0, 0}
\definecolor{FroestGreen}{rgb}{0.13,0.54,0.13}
\definecolor{tablegray}{gray}{.9}
\definecolor{codegreen}{rgb}{0,0.6,0}
\definecolor{codegray}{rgb}{0.5,0.5,0.5}
\definecolor{codepurple}{rgb}{0.58,0,0.82}
\definecolor{backcolour}{rgb}{0.95,0.95,0.92}
\fancypagestyle{plain}{%
\fancyhf{} 
\fancyfoot[C]{\sffamily\fontsize{9pt}{9pt}\selectfont\thepage} 

}
    \makeatletter
\def\@fnsymbol#1{\ensuremath{\ifcase#1\or $\ding{41}$ \or \dagger\or \ddagger\or
  \mathsection\or \mathparagraph\or \|\or **\or \dagger\dagger
  \or \ddagger\ddagger \else\@ctrerr\fi}}
    \makeatother

\usepackage{amsmath}
\begin{document}
\fancyhead{}
\newcommand{\bheading}[1]{{\vspace{4pt}\noindent{\textbf{#1}}}}
\newcolumntype{?}{!{\vrule width 1pt}}

\newcommand{\draft}[1]{\textcolor{blue}{#1}}
\newcounter{note}[section]
\renewcommand{\thenote}{\thesection.\arabic{note}}
\newcommand{\yz}[1]{\refstepcounter{note}{\bf\textcolor{red}{$\ll$YZ~\thenote: {\sf #1}$\gg$}}}
\newcommand{\mengyuan}[1]{\refstepcounter{note}{\bf\textcolor{blue}{$\ll$mengyuan~\thenote: {\sf #1}$\gg$}}}
\newcommand{\ZQ}[1]{\refstepcounter{note}{\bf\textcolor{red}{$\ll$ZL~\thenote: {\sf #1}$\gg$}}}
\newcommand{\tabincell}[2]{\begin{tabular}{@{}#1@{}}#2\end{tabular}}
\newcommand{\rebut}[1]{\textcolor{blue}{\bf Response: #1}}
\newcommand{\newexp}[1]{\textcolor{red}{\bf Response: #1}}

\newcommand{\figurewidth}{\columnwidth}
\newcommand{\secref}[1]{\mbox{Section~\ref{#1}}\xspace}
\newcommand{\secrefs}[2]{\mbox{Section~\ref{#1}--\ref{#2}}\xspace}
\newcommand{\figref}[1]{\mbox{Figure~\ref{#1}}}
\newcommand{\algref}[1]{\mbox{Algorithm~\ref{#1}}}
\newcommand{\tabref}[1]{\mbox{Table~\ref{#1}}}
\newcommand{\appref}[1]{\mbox{Appendix~\ref{#1}}}
\newcommand{\ignore}[1]{}

\newcommand{\etc}{\textit{etc.}\xspace}
\newcommand{\ie}{\textit{i.e.}\xspace}
\newcommand{\eg}{\textit{e.g.}\xspace}
\newcommand{\aka}{\textit{a.k.a.}\xspace}
\newcommand{\etal}{\textit{et al.}\xspace}

\newcommand{\atkname}{\textsc{CrossLine}\xspace}
\newcommand{\atknameOne}{\textsc{CrossLine V1}\xspace}
\newcommand{\atknameTwo}{\textsc{CrossLine V2}\xspace}
\newcommand{\atknameThree}{\textsc{CrossLine V3}\xspace}

\newcommand{\rip}{\textsc{RIP}\xspace}
\newcommand{\nrip}{\textsc{nRIP}\xspace}
\newcommand{\nextRIP}{\textsc{next\_rip}\xspace}
\newcommand{\vaddrInst}{\ensuremath{VA_0}\xspace}
\newcommand{\paddrInst}{\ensuremath{\textsc{PA}_I}\xspace}
\newcommand{\gcrInst}{\ensuremath{\textsc{gCR3}_I}\xspace}
\newcommand{\vaddrTarget}{\ensuremath{\textsc{VA}_T}\xspace}

\newcommand{\pageSPFN}{\ensuremath{\textsc{sPFN}_0}\xspace}
\newcommand{\pageGPFN}{\ensuremath{\textsc{gPFN}_0}\xspace}

\newcommand{\ctlTLB}{\textls[-50]{\textsc{TLB\_CONTROL\_FLUSH\_ALL\_ASID}}\xspace}

\newcommand{\gbytes}{\ensuremath{\mathrm{GB}}\xspace}
\newcommand{\mbytes}{\ensuremath{\mathrm{MB}}\xspace}
\newcommand{\kbytes}{\ensuremath{\mathrm{KB}}\xspace}
\newcommand{\bytes}{\ensuremath{\mathrm{B}}\xspace}
\newcommand{\hertz}{\ensuremath{\mathrm{Hz}}\xspace}
\newcommand{\ghertz}{\ensuremath{\mathrm{GHz}}\xspace}
\newcommand{\msecs}{\ensuremath{\mathrm{ms}}\xspace}
\newcommand{\usecs}{\ensuremath{\mathrm{\mu{}s}}\xspace}
\newcommand{\nsecs}{\ensuremath{\mathrm{ns}}\xspace}
\newcommand{\secs}{\ensuremath{\mathrm{s}}\xspace}
\newcommand{\gbits}{\ensuremath{\mathrm{Gb}}\xspace}

\newcounter{packednmbr}
\newenvironment{packedenumerate}{
\begin{list}{\thepackednmbr.}{\usecounter{packednmbr}
\setlength{\itemsep}{0pt}
\addtolength{\labelwidth}{4pt}
\setlength{\leftmargin}{12pt}
\setlength{\listparindent}{\parindent}
\setlength{\parsep}{3pt}
\setlength{\topsep}{3pt}}}{\end{list}}

\newenvironment{packeditemize}{
\begin{list}{$\bullet$}{
\setlength{\labelwidth}{0pt}
\setlength{\itemsep}{2pt}
\setlength{\leftmargin}{\labelwidth}
\addtolength{\leftmargin}{\labelsep}
\setlength{\parindent}{0pt}
\setlength{\listparindent}{\parindent}
\setlength{\parsep}{1pt}
\setlength{\topsep}{1pt}}}{\end{list}}

\makeatletter
\renewcommand\subsubsection{\@startsection{subsubsection}{3}{\z@}%
                       {-8\p@ \@plus -4\p@ \@minus -4\p@}
                       {-0.5em \@plus -0.22em \@minus -0.1em}%
                       {\normalfont\normalsize\bfseries\boldmath}}
\makeatother

\title{\atkname: Breaking ``Security-by-Crash'' based Memory Isolation in AMD SEV}




\author{Mengyuan Li}
\affiliation{%
  \institution{ The Ohio State University} \country{}}
\email{li.7533@osu.edu}
\author{ Yinqian Zhang}
\authornote{Corresponding authors}
\affiliation{%
  \institution{ Southern University of Science \& Technology} \country{}}
\email{yinqianz@acm.org}
\author{ Zhiqiang Lin}
\affiliation{%
  \institution{ The Ohio State University} \country{}}
\email{zlin@cse.ohio-state.edu}
\begin{abstract}

AMD's Secure Encrypted Virtualization (SEV) is an emerging security feature of modern AMD processors that allows virtual machines to run with encrypted memory and perform confidential computing even with an untrusted hypervisor. This paper first demystifies SEV's improper use of address space identifier (ASID) for controlling accesses of a VM to encrypted memory pages, cache lines, and TLB entries. We then present the \atkname attacks\footnote{\atkname refers to interference between telecommunication signals in adjacent circuits that causes signals to cross over each other.}, a novel class of attacks against SEV that allow the adversary to launch an attacker VM and change its ASID to that of the victim VM to impersonate the victim. We present two variants of \atkname attacks: \atknameOne decrypts victim's page tables or any memory blocks conforming to the format of a page table entry; \atknameTwo constructs encryption and decryption oracles by executing instructions of the victim VM. We discuss the applicability of \atkname attacks on AMD's SEV, SEV-ES, and SEV-SNP processors.



\end{abstract}

\ccsdesc[500]{Security and privacy~Hardware security implementation}
\ccsdesc[500]{Security and privacy~Hardware attacks and countermeasures}
\ccsdesc[500]{Security and privacy~Trusted computing}
\keywords{Trusted execution environments; Secure Encrypted Virtualization; Memory encryption; Cloud security} 
\maketitle

\thispagestyle{empty}
\pagestyle{empty}

\section{Introduction}

AMD's Secure Encrypted Virtualization (SEV) is a security extension for the AMD Virtualization (AMD-V) architecture~\cite{amd:2019:manual}, 
which allows one physical server to efficiently run multiple guest virtual machines (VM) concurrently on encrypted memory. When SEV is enabled, the memory pages used by a guest VM are transparently encrypted by a secure co-processor using an ephemeral key that is unique to each VM, thus allowing the guest VMs to compute on encrypted memory. SEV is AMD's ambitious movement towards confidential cloud computing, which is gaining traction in the cloud industry \cite{google:2020:sev}. 
Unlike traditional security assumptions in which the trustworthiness of the system software is taken for granted, SEV is built atop a threat model where system software including hypervisor can be untrusted. 

 \vspace{4pt}

\begin{center}
\minibox[rule=1pt,pad=2pt]{
\begin{minipage}[t]{0.9\columnwidth}
{\it ``SEV technology is built around a threat model where an attacker is assumed to have access to not only execute user level privileged code on the target machine, but can potentially execute malware at the higher privileged hypervisor level as well.''}~\cite{kaplan:2016:sevWpaper}.
\end{minipage}
}
\end{center}
\vspace{4pt}

Consequently, such an audacious threat assumption has been examined under the microscope with numerous attacks 
(e.g.,~\cite{hetzelt:2017:security, du:2017:sevUnsecure, buhren:2017:fault, morbitzer:2018:severed, Morbitzer:2019:extract, Li:2019:sevio, werner:2019:severest}) 
since its debut in 2017. With the assumption of a malicious hypervisor, these attacks successfully compromise the confidentiality and/or integrity provided by SEV's memory encryption by exploiting a number of design flaws, including unencrypted virtual machine control blocks (VMCB)~\cite{hetzelt:2017:security, werner:2019:severest}, unauthenticated memory encryption~\cite{hetzelt:2017:security, du:2017:sevUnsecure, buhren:2017:fault, Li:2019:sevio}, insecure ECB mode of memory encryption~\cite{du:2017:sevUnsecure, Li:2019:sevio}, unprotected nested page tables~\cite{morbitzer:2018:severed, Morbitzer:2019:extract}, and unprotected I/O operations~\cite{Li:2019:sevio}. 

In light of these security issues, AMD has enhanced SEV with a sequence of microcode and hardware updates, most notably SEV with Encrypted State (SEV-ES) and SEV with Secure Nested Paging (SEV-SNP). SEV-ES encrypts the VMCB of a VM to protect register values at VMEXITs; SEV-ES processors are already commercially available. 
To address the most commonly exploited flaw---the lack of memory integrity for SEV VMs (including unauthenticated memory encryption and unprotected nested page tables), AMD plans to release SEV-SNP, which introduces a Reverse Map Table (RMP) to dictate ownership of the memory pages, so that the majority of the previously known attacks will be mitigated.



However, in this paper, we move our attention to another, yet-to-be-reported design flaw of SEV---the improper ASID-based memory isolation and access control. Specifically, SEV adopts an ASID-based access control for guest VMs' accesses to SEV processor's internal caches and the encrypted physical memory. At launch time, each SEV VM is assigned a unique ASID, which is used as the tag of cache lines and translation lookaside buffer (TLB) entries. A secure processor (dubbed AMD-SP) that is in charge of generating and maintaining the ephemeral memory encryption keys also uses the current VM's ASID to index the keys for encrypting/decrypting memory pages upon memory access requests. As such, the ASID of an SEV VM plays a critical role in controlling its accesses to the private data in the cache-memory hierarchy. Nevertheless, the assignment of ASID to a VM is under complete control of the hypervisor. An implicit  ``security-by-crash'' security principle is adopted in the SEV design:
\vspace{2pt}
\begin{center}
\minibox[rule=1pt,pad=2pt]{
\begin{minipage}[h]{0.9\columnwidth}
{\it ``Although the hypervisor has control over the ASID used to run a VM and select the encryption key, this is not considered a security concern since a loaded encryption key is meaningless unless the guest was already encrypted with that key. If the incorrect key is ever loaded or the wrong ASID is used for a guest, the first instruction fetch of that guest will fail as memory will be decrypted with the wrong key, causing junk data to be executed (and very likely causing a fault).''}~\cite{kaplan:2016:sevWpaper}
\end{minipage}
}
\end{center}
\vspace{2pt}


The aim of this paper, therefore, is to investigate the validity of this ``security-by-crash'' design principle. To do so, we first study how ASIDs are used in SEV processors to isolate encrypted memory pages, CPU caches, and TLBs. We also explore how ASIDs are managed by the hypervisor, how an ASID of a VM can be altered by the hypervisor at runtime, and why the VM with altered ASID crashes afterwards. This exploration leads to the discovery of several potential opportunities for a VM with an altered ASID to \textit{momentarily} breach the ASID-based memory isolation before it crashes.  \looseness=-1

Next, based on our exploration, we then present \atkname attacks, which exploit such a \textit{momentary execution} to breach the confidentiality and integrity of SEV VMs. Specifically, an adversary controlling the hypervisor can launch an attacker VM and, during its VMEXIT, assign it with the same ASID as the victim VM, and then resume it, leading to the violation of the ASID-based access control to the victim's encrypted memory. 

We mainly present two variants of \atkname. In \atknameOne, even though no instructions are executed by the attacker VM after VMRUN, we show that it is possible to load memory pages encrypted with the victim VM's memory encryption key (VEK) during page table walks, thus revealing the encrypted content of the ``page table entries'' (PTE) through nested page faults. This attack variant enables the adversary to extract the entire encrypted page table of the SEV guest VM, as well as any memory blocks conforming to the PTE format. We have also successfully demonstrated \atknameOne on SEV-ES machines, in which we devise techniques to bypass the integrity checks of launching the attacker VM with the victim VM's encrypted VMCB, while keeping the victim VM completely unaffected. 
In \atknameTwo, by carefully crafting its nested page tables, the attacker VM could manage to momentarily execute arbitrary instructions of the victim VM. By wisely selecting the target instructions, the adversary is able to construct encryption oracles and decryption oracles, which enable herself to breach both integrity and confidentiality of the victim VM. \atknameTwo is confined by SEV-ES, but its capability is stronger than V1.





\bheading{Differences from known attacks.}
\atkname differs from all previously demonstrated SEV attacks in several aspects. \textit{First}, \atkname does not rely on SEV's memory integrity flaws, which is a common pre-requisite for all known attacks on SEV. Although \atkname may not work on SEV-SNP, the protection does not come from memory integrity, but a side-effect of the RMP implementation. \textit{Second}, \atkname attacks do not directly interact with the victim VMs and thus enable \textit{stealthy} attacks. As long as the ephemeral encryption key of the victim VM is kept in the AMD-SP and the victim's encrypted memory pages are not deallocated, \atkname attacks can be performed even when the victim VM is shutdown. Therefore, \atkname is undetectable by the victim VM. In contrast, prior attacks relying on I/O operations of the victim VM~\cite{Li:2019:sevio, du:2017:sevUnsecure, morbitzer:2018:severed, Morbitzer:2019:extract} are detectable by the victim VM.

\atkname attacks question a fundamental ``security-by-crash'' security principle underpinning the design of SEV's memory and cache isolation. The demonstration of \atkname suggests that SEV should not rely on adversary-controlled ASIDs to mediate access to the encrypted memory. 
To eliminate the threats, a principled solution is to maintain the identity of VMs in the hardware, which unfortunately requires some fundamental changes in the architecture. As far as we know, SEV-SNP will not integrate such changes. 
\looseness=-1

\bheading{Responsible disclosure.} 
We have disclosed \atkname attacks to AMD via emails in December 2019 and discussed the paper with AMD engineers by phone in January 2020. 
We have pointed out several vulnerable hardware designs, including: (1) The lack of ASID authentication and inappropriate “security-by-crash” principle; (2) the lack of triple fault reporting, which allows SEV and SEV-ES VM to resume from a triple fault by rewinding VMCB; (3) the VMSA check is only tied to VMSA's physical address but not VMCB's physical address, which makes Crossline work in SEV-ES. These vulnerabilities have been acknowledged by AMD.
The demonstrated attacks and their novelty have been acknowledged. As discussed in the paper, neither of the two attack variants directly affect SEV-SNP. Therefore, AMD would not replace ASID-based isolation in the short term, but may invest more principled isolation mechanisms in the future. 



\bheading{Contributions.} This paper makes the following contributions to the security of AMD SEV and other trusted execution environments. 

\begin{packeditemize}
\item It investigates SEV's ASID-based memory, cache, and TLB isolation, and demystifies its ``security-by-crash'' design principle (\S\ref{sec:understand}). It raises security concerns of the ``security-by-crash'' based memory and TLB isolation for the first time.

\item It presents two variants of \atkname attacks---the only attacks that breach the confidentiality and integrity of an SEV VM without exploiting SEV's memory integrity flaws (\S\ref{sec:attacks}). 

\item It presents successful attacks against SEV and SEV-ES processors (\S\ref{sec:countermeasure}). It also discusses the applicability of \atkname on the upcoming SEV-SNP processors (\S\ref{sec:discuss}).

\end{packeditemize}


\section{Background}
\label{sec:bg}


\bheading{Secure Memory Encryption (SME).}
SME is AMD's x86 extension for real-time main memory encryption, which is supported in AMD CPU with Zen micro architecture from 2017~\cite{singh:2017:zen}. Aiming to defeat cold boot attack and DRAM interface snooping, an embedded Advanced Encryption Standard (AES) engine encrypts data when the processor writes to the DRAM and decrypts it when processor reads it. The entire DRAM is encrypted with a single ephemeral key which is randomly generated each time the machine is booted. A 32-bit ARM Cortex-A5 Secure Processor (AMD-SP)~\cite{lai:2013:sever_security} is integrated in the system-on-chip (SOC) alongside the main processor, providing a dedicated security subsystem, storing, and managing the ephemeral key. Although all memory pages are encrypted by default, the operating system can mark some pages as unencrypted by clearing the \textit{C-bit} (the 48th bit) of the corresponding page table entries (PTE). 
However, regardless of the C-bit, all code pages and page table pages are encrypted by default. With Transparent SME (TSME), a special mode of operation of SME, the entire memory is encrypted, ignoring the C-bits of the PTEs. 


\bheading{AMD Virtualization (AMD-V).}
AMD-V is a set of extensions of AMD processors to support virtualization. Nested Page Tables (nPT) is introduced by AMD-V to facilitate address translation~\cite{amd:2008:npt}. 
AMD-V's nPT provides two levels of address translation. When nPT is enabled, the guest VM and the hypervisor have their own CR3s: a guest CR3 (gCR3) and a nested CR3 (nCR3). The gCR3 contains the guest physical address of the guest page table (gPT); the nCR3 contains the system physical address of the nPT. 
To translate a virtual address (gVA) used by the guest VM into the system physical address (sPA), the processor first references the gPT to obtain the guest physical address (gPA) of each page-table page. To translate the gPA of each page, an nPT walk is performed. During a nPT walk, the gPA is treated as host virtual address (hVA) and translated into the sPA using the nPT.  


Translation lookaside buffers (TLB) and Page Walk Cache (PWC) are internal buffers in AMD processors for speeding up the address translation. AMD-V also relies on these internal buffers for performance improvements. 
AMD-V further introduces an nTLB for nPT. A successful nPT walk caches the translation from gPA to sPA in the nTLB for fast accesses~\cite{amd:2019:manual}, while the normal TLBs are used to store translations from virtual addresses of either the host or the guest to sPA. 
\looseness=-1

To exchange data between the hypervisor and the guest VMs, a data structure dubbed the virtual machine control block (VMCB) is located on a shared memory page. VMCB stores the guest VM's register values and some control bits during VMEXIT. The VMCB is under the control of the hypervisor to configure the behaviors of the guest VM.


\bheading{Secure Encrypted Virtualization (SEV).} SEV combines AMD-V architecture with SME to allow individual VMs to have their own VM Encryption Key (VEK)~\cite{amd:2019:sevapi}. Each VEK is generated by the processor and assigned to an SEV VM when launched by the hypervisor. All VEKs are stored in the AMD-SP and are never exposed to DRAM during their entire life cycle. SEV distinguishes different VEKs using ASIDs. 
When a memory request is made, the AMD-SP determines which key to be used with the current ASID, 
achieving page-granular memory encryption with different keys.  

\section{Demystifying ASID-based Isolation}
\label{sec:understand}

\begin{figure}[t]
\centering
\includegraphics[width=0.99\columnwidth]{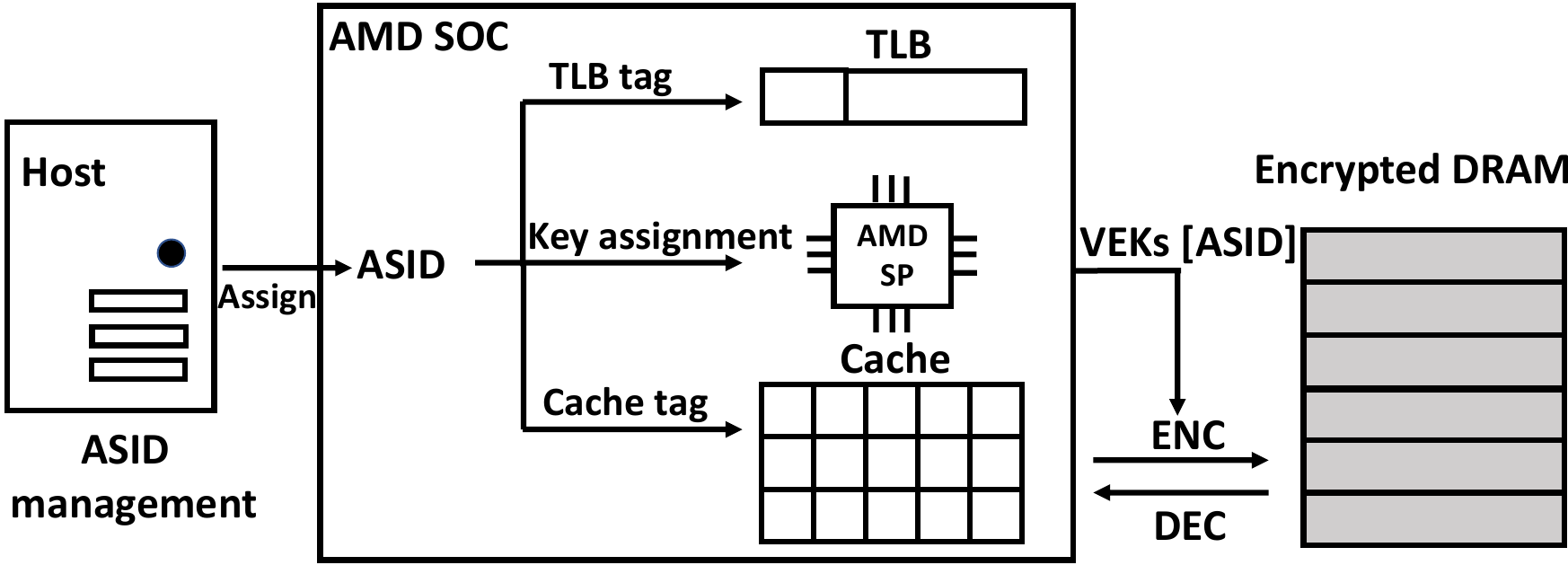}
\caption{ASID-based memory isolation in SEV.}
\label{fig:ASID}
\end{figure}

ASID was initially designed by AMD to tag TLB entries so that unnecessary  TLB flushes can be avoided when switching between guest VMs and the host. SEV reuses ASID as the indices of VEKs stored in AMD-SP. Cache tags are also extended accordingly to isolate cache lines with different ASIDs. As a result, ASID becomes the de-facto identifier used by SEV processors to control the software's accesses to virtual memory, caches, and TLBs (\figref{fig:ASID}). 

However, following AMD-V, SEV allows the hypervisor to have (almost) complete authority over the management of ASIDs, 
which gives rise to security concerns as a malicious hypervisor may abuse this capability to breach ASID-based isolation.  
Interestingly, AMD adopts a ``security-by-crash'' principle and assumes if ``{\it the wrong ASID is used for a guest}'', the execution of the instruction will ``{\it likely cause a fault}''~\cite{kaplan:2016:sevWpaper}. In this section, we set off to understand and demystify how ASIDs are used to isolate memory, cache, and TLBs in SEV, and how ASIDs are managed by the hypervisor.


\subsection{ASID-based Isolation}


\subsubsection{\bf ASID-based Memory Isolation}

ASIDs are used by the AMD-SP to index VEKs of SEV VMs. The SEV hardware ensures the data and code of an SEV VM is encrypted in the DRAM and only decrypted when loaded into the SOC. Specifically, each memory read from an SEV VM consists of memory fetches by the memory controller of a 128-bit aligned memory block, followed by an AES decryption by AMD-SP using the VEK corresponding to the current ASID. The current ASID is an integer stored in a hidden register of the current CPU core, which cannot be accessed by software in the guest VM. 
\looseness=-1

SEV allows the guest OS to decide, by setting or clearing the C-bit of the PTE, whether each virtual memory page is (treated as) private (encrypted with the guest's VEK) or shared (either encrypted with the host's VEK or unencrypted). For instance, when the C-bit of a page is set, memory reads from this virtual-physical mapping is considered encrypted with the guest VM's VEK, regardless of its true encryption state, and thus a memory read in that page will be decrypted using the VEK of the current ASID. By default, the guest VM sets the guest C-bits for private pages during the boot period. 


However, the hypervisor is able to manipulate the nested C-bit (nC-bit) in nPT. When the gC-bit (the C-bit of the gPT) conflicts with the nC-bit, AMD-SP encrypts the memory pages according to the following rules: When gC-bit=0 and nC-bit=1, the page is encrypted with the hypervisor's VEK; when gC-bit=1, regardless of the nC-bit, the page is encrypted with the guest VM's VEK; when gC-bit=0 and nC-bit=0, the page is not encrypted. Following SME, the code pages are always considered private to the guest VM and thus is always encrypted regardless of the guest C-bits. Similarly, the gPT is also always encrypted with the guest's VEK. 

\subsubsection{\bf ASID-based TLB Isolation}

ASID was originally introduced to avoid TLB flushes when the execution context switches between the guest VM and the hypervisor, which is achieved by extending each TLB tag with ASID. With the ASID capability, when observing activities like MOV-to-CR3, context switches, updates of {\tt CR0.PG/CR4.PGE/CR4.PAE/CR4.PSE}, the hardware does not need to flush the entire TLB, but only the TLB entries tagged with the current ASID~\cite{amd:2019:manual}. However, to properly isolate TLB, the management of ASIDs for non-SEV VMs and SEV VMs is slightly different. 


\bheading{Non-SEV VMs.}
Each VCPU of a non-SEV VM may have different ASIDs, which can be assigned dynamically before each VMRUN. More specifically, before the hypervisor is about to resume a VCPU with VMRUN, it checks if the VCPU was the one running on this CPU core before the control was trapped into the hypervisor. If so, the hypervisor keeps the ASID of the VCPU unchanged and resumes the VCPU directly; if not, the hypervisor selects another ASID (from the ASID pool) and assign it to the VCPU. In the former case, TLB entries can be reused by the VCPU as its ASID is unchanged. However, in the latter case, the residual TLB entries (tagged with ASID of the hypervisor or the previous VCPU) should not be reused. 


\bheading{SEV VMs.} SEV processors rely on a similar strategy to isolate entries in the TLBs with ASID. However, instead of dynamically assigning an ASID to a VCPU before VMRUN, all VCPUs of the same SEV VM are assigned the same ASID at launch time, which remains the same during the entire life cycle of the SEV VM.
%




\subsubsection{\bf ASID-based Cache Isolation}

On platforms that support SEV, cache lines are tagged with the VM's ASID indicating to which VM this data belongs, thus preventing the data from being misused by entities other than its owner~\cite{kaplan:2016:sevWpaper}. When data is loaded into cache lines, according to the current ASID, AMD-SP automatically decrypts the data with the corresponding VEK and stores the ASID value into the cache tag. When a cache line is flushed or evicted, AMD-SP uses the ASID in the cache tag to determine which VEK to use when encrypting this cache line before writing it back to DRAM. The cache tag is also extended to include the C-bit~\cite{kaplan:2016:sevWpaper}. Because the cache is now tagged with ASID and C-bit, cache coherence of the same physical address is not maintained if the two virtual memory pages do not have the same ASID and C-bit.  \looseness=-1

\ignore{
\begin{figure}[t]
\centering
\begin{subfigure}[b]{0.49\columnwidth}
\includegraphics[width=\textwidth]{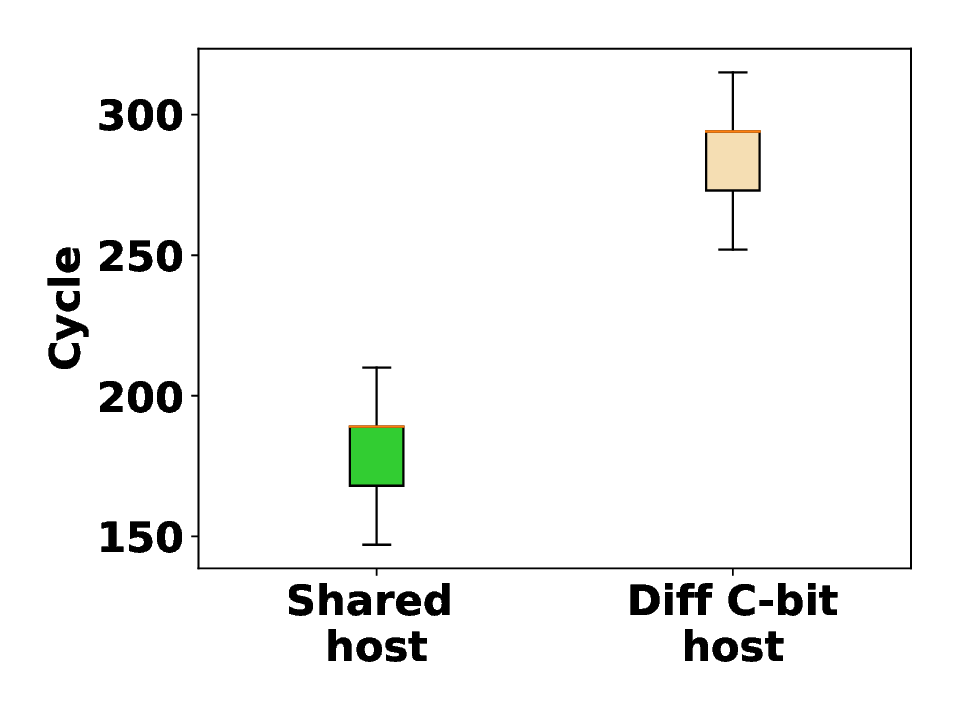}
\caption{host/host}
\label{fig:flush_reload_host}
\end{subfigure}
\hfill
\begin{subfigure}[b]{0.49\columnwidth}
\includegraphics[width=\textwidth]{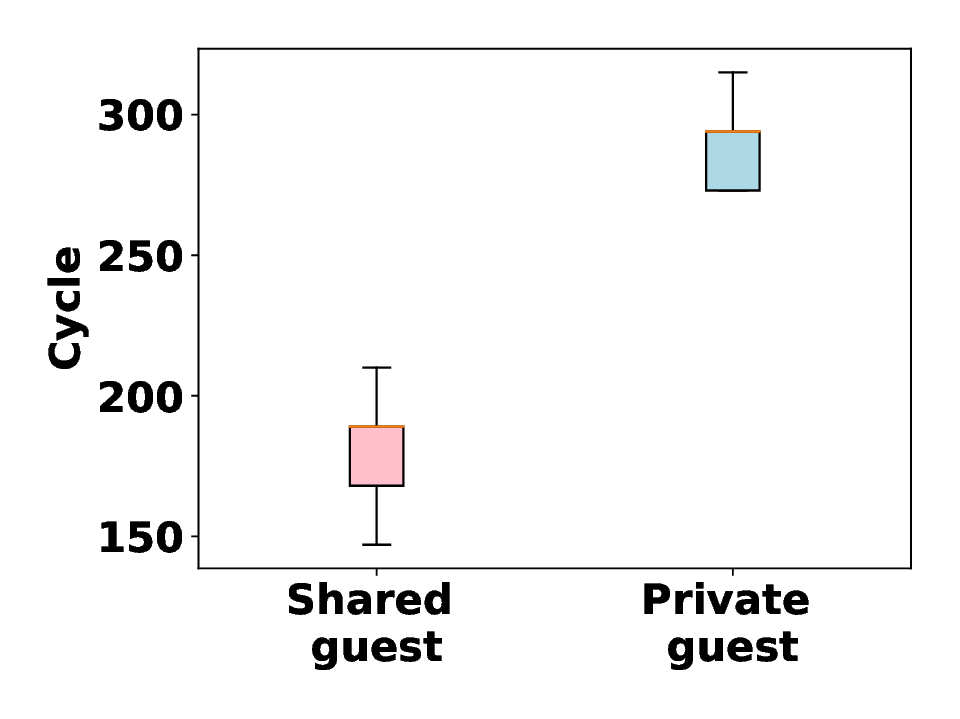}
\caption{Guest/host}
\label{fig:flush_reload_guest}
\end{subfigure}
\caption{Flush-reload.}
\label{fig:flushreload}
\vspace{-10pt}
\end{figure}
}




\ignore{
\begin{figure}[t]
\centering
\includegraphics[width=\columnwidth]{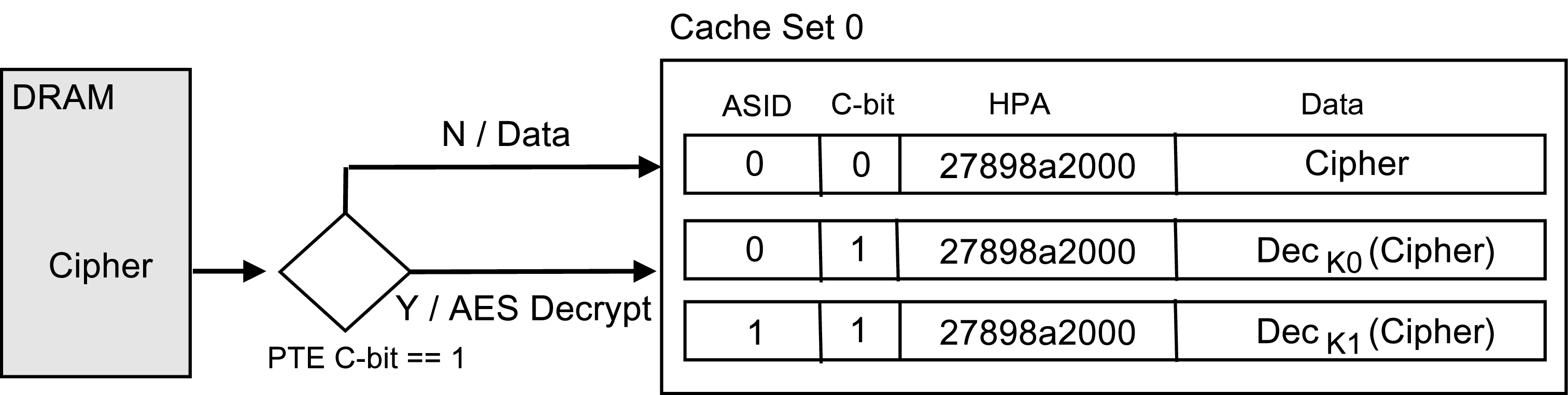}
\caption{Multi-copy-cache.}
\label{fig:Multi-copy-cache}
\end{figure}
}

\subsection{ASID Management}



\subsubsection{\bf ASID Life Cycle}

The hypervisor reserves a pool (\ie, a range of integers) of available ASIDs for all VMs (we call all-ASID pool for simplicity), and a separate pool of ASIDs for SEV VMs (SEV-ASID pool). The maximum ID number of the all-ASID pool is determined by CPUID 0x8000000a[EBX] (\eg, 32768, thus the available ASIDs are whole numbers between 1 and 32767). The maximum ID number of the SEV-ASID pool is determined by CPUID 0x8000001f[ECX] (\eg, 15, which suggests the legal ASIDs for SEV VMs are 1 to 15). Note that ASID 0 is reserved for the host OS (i.e., hypervisor), and is also not  allowed to be assigned to a VCPU for processors with or without SEV extensions~\cite{amd:2019:manual}.


On SEV platforms, the hypervisor uses {\tt ACTIVATE} command to inform AMD-SP that a given guest is bound with an ASID and uses {\tt DEACTIVATE} command to de-allocate an ASID from the guest. {\tt DEACTIVATE} also uninstalls the guest VM's VEK. After a successful {\tt DEACTIVATE}, if there is no available ASID in the SEV-ASID pool, the hypervisor may re-allocate the ASID to another VM~\cite{amd:2019:sevapi}.

At runtime, when the processor runs under the guest mode, the guest VM's ASID is stored in the ASID register that is hidden from software; when the processor runs under the host mode, the register is set to 0, which is the hypervisor's ASID. The guest VM's ASID is stored at the VMCB during VMEXIT. After VMRUN the processor restores the ASID in the VMCB. 
The VMCB State Cache allows the processor to cache some guest register values between VMEXIT and VMRUN for performance enhancement. The physical address of the VMCB is used to perform access control of the VMCB State Cache. However, the VMCB clean field controlled by the hypervisor can be used to force the processor to discard selected cached values. For example, bit-$2$ of the VMCB clean field indicates that an ASID reload is needed; bit-$4$ of the clean field indicates fields related to nest pages are dirty and needed to be reloaded from the VMCB. Some VMCB fields are strictly not cached and the corresponding register values will be reloaded from the VMCB every time. For example, offset $058$h of the VMCB is a TLB control field to indicate whether the hardware needs to flush TLB after VMRUN; this field is always uncached.

\subsubsection{\bf ASID Restrictions} SEV implements both launch-time and run-time restrictions about ASID.

\bheading{Launch-time restrictions.} On processors supporting SEV, the hypervisor cannot bind a current active ASID in the SEV-ASID pool to an SEV VM during launch time~\cite{amd:2019:sevapi}. 
However, an adversary is able to deactivate the victim SEV VM and then activate an attacker SEV VM with the same ASID. The hardware requires the hypervisor to execute a {\tt WBINVD} instruction and a {\tt DF\_FLUSH} instruction after deactivating an ASID and before re-activating it. The {\tt WBINVD} flushes all modified cache lines and invalidates all cache lines. The {\tt DF\_FLUSH} instruction flushes data fabric write buffers of all CPU cores. If these instructions are not executed before associating the ASID with a new VM, a {\tt WBINVD\_REQUIRED} or {\tt DFFLUSH\_REQUIRED} error will be returned by the AMD-SP and the VM launch process will be terminated. \looseness=-1

This restriction is critical to the isolation of cache lines. Otherwise, victim VM's residual cache data can be read by subsequent attacker VM. In particular, the attacker VM can use the {\tt WBINVD} instruction to flush the cache data to memory. Cache lines belonging to victim VM will thus be encrypted with the attacker VM's VEK and then flushed into the memory. Subsequent reads to those memory data will return plaintext and thus allow the adversary to extract the data. 


\bheading{Run-time restrictions.}
After a VM is launched, the hypervisor can change its ASID during VMEXITs, by changing the ASID field of its VMCB, which will take effect when the VM is resumed. There is no additional hardware restriction at runtime. 
As such, it is possible to have two SEV VMs with the same ASID on the same machine, though the one with an incorrect ASID will crash very soon. 

Moreover, the VMCB also contains a field ($090$h) to indicate if the VM is an SEV VM or a non-SEV VM. Therefore, it is possible to first launch an SEV VM and a non-SEV VM with the same ASID, and then, during VMEXITs of the non-SEV VM, change the non-SEV VM into an SEV VM by setting the corresponding bit in the VMCB. We have experimentally confirmed this possibility on our testbed (as shown in \secref{sec:discuss:v3}). It suggests that the hardware trusts the values of VMCB to determine (1) if the VM to be resumed is an SEV VM and (2) what ASID is associated with it. The hardware does not store this information to a secure memory region and use it for validation. The only additional validation performed by the AMD-SP is that the ASIDs of SEV VMs must fall into the valid ranges\footnote{The lower portion of the valid ASID range of SEV VMs are reserved for SEV-ES VMs. CPUID Fn8000\_001F[ECX] specifies valid SEV ASIDs and CPUID Fn8000\_001F[EDX] specifies the minimum ASID values used for SEV (but non-SEV-ES) VMs.}.
Therefore, while a VM was launched as a non-SEV VM, we can effectively (though momentarily) make it an SEV VM with the same ASID as another SEV VM.

\subsubsection{\bf ``Security-by-Crash''}

As the hypervisor has the liberty of changing the ASIDs of both SEV VMs and non-SEV VMs, security concerns arise when the hypervisor is not considered a trusted party. However, AMD believes that when an SEV VM is resumed with an ASID different from its own, its subsequent execution will lead to unpredictable results and eventually crash the VM~\cite{kaplan:2016:sevWpaper}. 

Specifically, to change the ASID of a VM (either an SEV or non-SEV VM), the hypervisor can directly edit the ASID field of the VMCB, set the VMCB clean-field to inform the hardware to bypass the VMCB State Cache, and then resume the VM with VMRUN. After the VM is resumed, if the {\tt RFLAGS.IF} bit in the VMCB is set, the virtual address specified by the interrupt descriptor-table register (IDTR) will be accessed, because the guest OS will try to handle interrupts immediately; if the {\tt RFLAGS.IF} bit is cleared, the instruction pointed to by \nrip---the next sequential instruction pointer---is going to be fetched and executed. However, in either case, the virtual address translation will cause problems. 

First, any TLB entries remaining due to its previous execution becomes invalid because its ASID has been changed; the ASID tag in the TLB entries would not match. Second, a page table walk is unlikely to succeed, as its own page tables are encrypted using the VEK indexed by its own ASID. As a result, the top-level page table will be decrypted into meaningless bit strings. References to a \textit{page table entry} of this page will trigger an exception to be handled by the guest OS. Finally, a handler of the guest OS is to be invoked to handle the exception. However, any reference of this handler will be decrypted using an incorrect VEK, leading to a \textit{triple fault} that eventually crashes the VM.

\subsection{Summary}


We highlight a few key points of SEV's ``security-by-crash'' based memory isolation mechanisms.

\begin{packeditemize}

\item \textbf{ASID is used for access control.} ASID is the only identifier used for controlling accesses to virtual memory, caches, and TLBs. Once a VM is successfully resumed from VMEXIT, the CPU and AMD-SP only rely on the ASID (loaded from its VMCB) to validate memory requests.


\item \textbf{ASID is managed by the hypervisor.} The hypervisor may assign any ASID (including the ASID of another active SEV VM) to an SEV or non-SEV VM during VMEXIT. The only restriction enforced by the hardware is that the ASID must fall into the range in accordance with the VM's SEV type. 

\item \textbf{Security is achieved by VM crash.} The security of the mechanism relies solely on the faults triggered during the execution of the VM if its ASID has been changed. The faults can be caused by memory decryption with an incorrect VEK during instruction fetches or page table walks.

\item \textbf{Cache/TLB entries are flushed by the hypervisor.} The hypervisor controls whether and when to flush TLB and cache entries associated with a specific ASID. Only limited constraints are enforced by the hardware during ASID activation. Misuse of these resources is possible.




\end{packeditemize}

\section{\atkname Attacks}
\label{sec:attacks}


The goal of our \atkname attacks is to extract the memory content of the victim VM that is encrypted with the victim VM's VEK. We make no assumption of the adversary's knowledge of the victim VM, including its kernel version, the applications running in it, \etc The common steps of the \atkname attacks are the following: (1) the adversary who controls the hypervisor launches a carefully crafted attacker VM; (2) 
the hypervisor alters the ASID of the attacker VM to be the same as that of the victim VM during VMEXITs; 
(3) the hypervisor prepares a desired execution environment for the attacker VM by altering its VMCB and/or its nPT; (4) the attacker VM resumes after VMRUN, allowing a \textit{momentary execution} before it crashes. During the momentary execution, memory accesses from the attacker VM will trigger memory decryption using the victim VM's VEK. 

Although the attacker VM crashes shortly---due to the ASID-based isolation in TLB, caches, and memory---we show that this momentary execution, though very brief, already enables the attacker VM to impersonate the victim VM and breach its confidentiality and integrity. 
Note that the only requirement of the victim VM at the time of the attack is that it has been launched and the targeted memory pages have been encrypted in the physical memory. Whether or not the victim VM is concurrently running during the attack is not important. Therefore, \atkname is stealthy in that it does not interact with the victim VM at all. Detection of such attacks from the victim VM itself is unlikely.


\subsection{Variant 1: Extracting Encrypted Memory through Page Table Walks}

\begin{figure}[t]
\centering
\includegraphics[width=0.95\columnwidth]{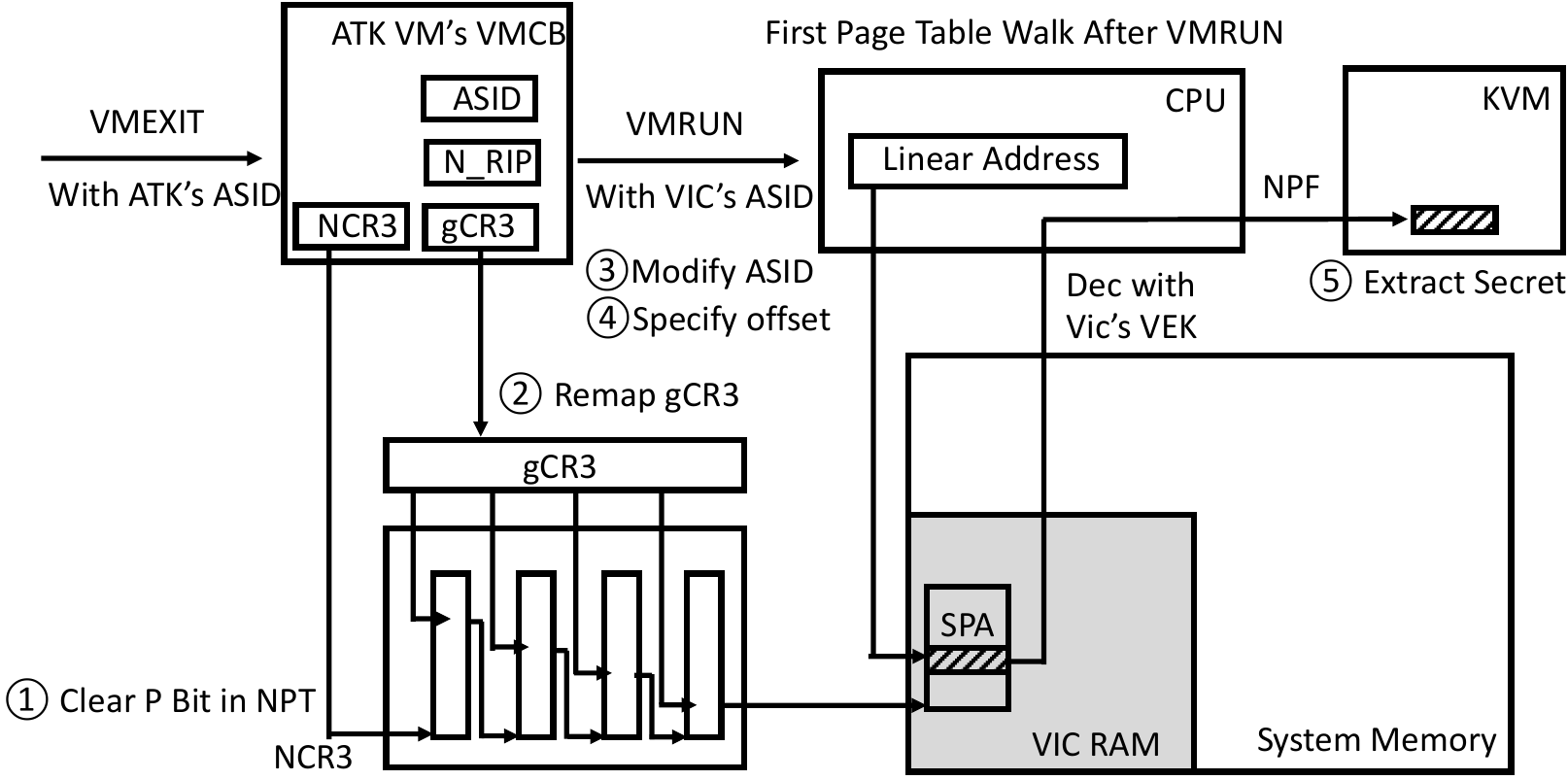}
\caption{Workflow of \atknameOne.}
\label{fig:workflow}
\end{figure}

The \atknameOne explores the use of nested page table walks during the momentary execution to decrypt the victim VM's memory protected by SEV. 
To ease the description, let the victim VM's ASID be 1 and the attacker VM's ASID be 2. We use \pageSPFN to denote the system page frame number of the targeted memory page encrypted with the victim VM's VEK. We use sPA$_0$ to denote the system physical address of one 8-byte aligned memory block on \pageSPFN, which is the target memory the adversary aims to read.   
The workflow of \atknameOne is shown in \figref{fig:workflow}. When the hypervisor handles a VMEXIT of the attacker VM, the following steps are executed: \looseness=-1



\bheading{\ding{192} Clear the Present bits.} 
The hypervisor alters the attacker VM's nPT to clear the Present bits of the PTEs of all memory pages. Thereafter, any memory access from the attacker VM after VMRUN will trigger a nested page fault, because the mapping from gPA to sPA in the nPT is missing. 


\bheading{\ding{193} Remap the current gCR3 of the attacker VM.} The hypervisor remaps the gCR3 of the current process in the attacker VM by altering the nPT. Now the gCR3 maps to \pageSPFN. The hypervisor then sets the Present bit of this new mapping in the nPT.

\bheading{\ding{194} Modify the attacker VM’s VMCB.} The hypervisor changes the attacker VM's ASID field in the VMCB to the victim VM's ASID (from 2 to 1 in this example).  

\bheading{\ding{195} Specify the targeted page offset.} Before resuming the attacker VM with VMRUN, the hypervisor also modifies the value of \nrip in VMCB to specify which offset (\ie, sPA$_0$) of the target page (\ie, \pageSPFN) to decrypt. Specifically, in a 64-bit Linux OS, bits 47 to 12 of a virtual address are used to index the page tables: bits 47-39 for the top-level page table; bits 38-30 for the second-level; bits 29-21 for the third; and bits 20-12 for the last-level page table. Each 4KB page in the page table has 512 entries (8 bytes each) and each entry contains the page frame number of the memory page of next-level page table or, in the case of the last-level page table, the page frame number of the target address. \atknameOne exploits the top-level page table walk to decrypt one 8-byte block each time. To control the offset of the 8-byte block within the page, the adversary modifies the value of \nrip stored in the VMCB so that its bit 47-39 can be used to index the top-level page table. The algorithm to choose \nrip  properly is specified in Algorithm~\ref{alg:offset}. Specifically, if the offset is less than 0x800, the \nrip is set to be in the range of 0x0000000000000000 - 0x00007fffffffffff (canonical virtual addresses of user space); if the offset is greater than or equal to 0x800, the \nrip is set to be in the range of 0xffff800000000000 - 0xffffffffffffffff (canonical virtual addresses of kernel space).

\bheading{\ding{196} Extract secrets from nested page faults.} 
After VMRUN, the resumed attacker VM immediately fetches the next instruction to be executed from the memory. This memory access is performed with ASID=1 (\ie, the victim VM's ASID). The address translation is also performed with the same ASID. As the TLB does not hold valid entries for address translation, and thus an address translation starts with a page table walk from the gCR3, which maps to \pageSPFN in the nPT. Therefore, an 8-byte memory block on \pageSPFN, whose offset is determined by bit 47-39 of the virtual address of the instruction, is loaded by the processor as if it is an entry of the page table directory. As long as the corresponding memory block conforms to the format of a PTE (to be described shortly), the data can be extracted and notified to the adversary as the faulting address (encoded in the EXITINFO2 field of VMCB).  


\begin{algorithm}[t]
{\footnotesize
 initialization\;
 \While{dumping the page}{
  try to dump 8-byte memory block sPA$_0$ \;
  \eIf{sPA$_0$\% 0x1000 $<$ 0x800}{
    \nrip = 0x8000000000* (sPA$_0$\% 0x1000 / 0x8)\; 
   }{
    \nrip = 0xffff000000000000 + 0x8000000000* (sPA$_0$\% 0x1000 / 0x8)\;
  }
 }
}
 \caption{\footnotesize Determine \nrip when dumping one layer of page table (4096 bytes)}
 \label{alg:offset}
\end{algorithm}

\subsubsection{\bf Dumping Victim Page Tables}

A direct security consequence of \atknameOne is to dump the victim VM's entire guest page table, which is deemed confidential as page-table pages are always encrypted in SEV VMs regardless of the C-bit in the PTEs.  \looseness=-1


To dump the page tables, the adversary first locates the root of the victim VM's guest page table specified by its gCR3. She can do so by monitoring the victim VM's page access sequence using page-fault side channels. Specifically, during the victim VM's VMEXIT, the adversary clears the Present bit of all page entries of the nPT of the victim VM, evicts all the TLB entries, invalidates the nPT entries cached by nTLB and PWC. After VMRUN, the victim VM immediately performs a page table walk. The gPA of the first page to be accessed is stored in its gCR3. The adversary thus learns the gPA of the root of the guest page table. Once each of the entries of the root page table is extracted by \atknameOne, the rest of the page table can be decrypted one level after another.

\bheading{Evaluation.}
We evaluated this attack on a blade server with an 8-Core AMD EPYC 7251 Processor. The host OS runs Ubuntu 64-bit 18.04 with Linux kernel v4.20 and the guest VMs run Ubuntu 64-bit 18.04 with Linux kernel v4.15 (SEV supported since v4.15). The QEMU version used was QEMU 2.12. 
The victim VMs were SEV-enabled VMs with 4 virtual CPUs, 4 GB DRAM and 30 GB disk storage. The attacker VMs were SEV-enabled VMs with only one virtual CPU, 2 GB DRAM and 30 GB disk storage. All the victim VMs were created by the ubuntu-18.04-desktop-amd64.iso image with no additional modification. 

After decrypting one 8-byte memory block, the attacker VM is trapped by a triple fault, which indicates the VM itself cannot handle the error. In order to continue decrypting other memory blocks, one intuitive solution is to reboot the attack VM every time there is a triple fault. Our empirical evaluation sugggests that it takes around 2 seconds to decrypt one 8-byte memory block (including a VM reboot). 
To speed up the memory decryption, the adversary could take the following \textit{VMCB rewinding} approach: After extracting one 8-byte block through a VMEXIT caused by the nested page fault, the adversary could continue to decrypt the next 8-byte block without rebooting the attacker VM. To do so, the adversary directly repeats the attack steps by rewinding the VMCB of the attacker VM to the previous state and changing the \nrip to perform the next round of attack. With this approach, we found the average time (over 500 trials) to decrypt a 4KB memory page by a single  attacker VM was only 39.580ms (with a standard deviation of 4.26ms).



\subsubsection{\bf Reading Arbitrary Memory Content}

Beyond page tables, the adversary could also extract regular memory pages of the victim VM. For example, if the data of an 8-bytes memory block is {\tt 0x00 0x00 0xf1 0x23 0x45 0x67 0x8e 0x7f}, the extracted data through page fault is {\tt 0x712345678}; if the data is {\tt 0x00 0x00 0x0a 0xbc 0xde 0xf1 0x20 0x01}, the extracted data is {\tt 0xabcdef12}. However, as \atknameOne only reveals the encrypted data as a page frame number embedded in the PTE, such memory decryption only works on 8-byte aligned memory blocks (\ie, the begin address of the block is a multiple of 8 and the size of the block is also 8 bytes) that conforms to the format of a PTE. 

\begin{figure}[t]
\centering
\includegraphics[width=0.95\columnwidth]{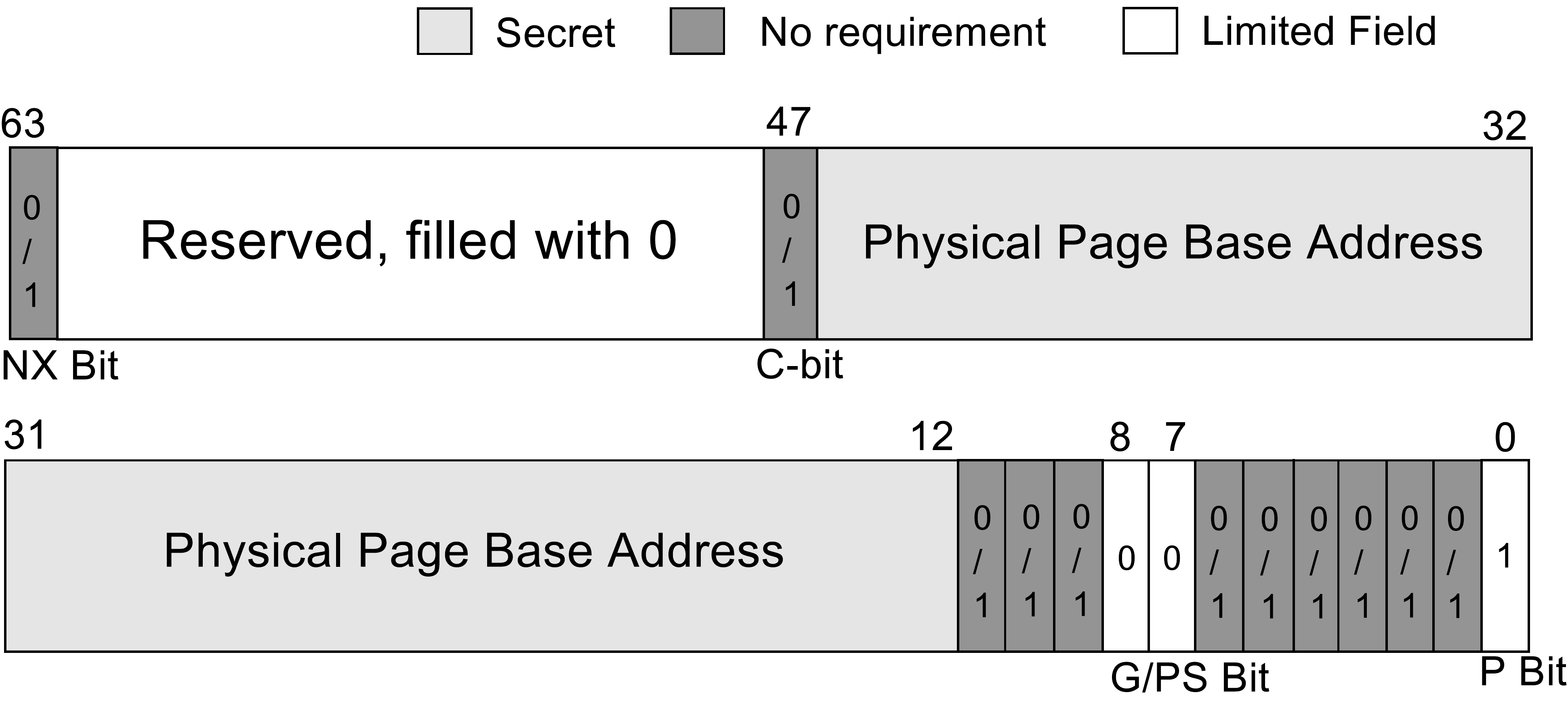}
\caption{Valid PTE format.}
\label{fig:boundary}
\end{figure}

Concretely, as shown in \figref{fig:boundary}, the 8-byte memory block to be extracted from \atkname, must satisfy the following requirements: The Present bit (bit 0) must be 1; Bits 48-62 must be all 0s, and Bits 7-8 are both 0s (optional).
\ignore{
\begin{packeditemize}
\item The Present bit (bit 0) must be 1.
\item Bits 48-62 must be all 0s.
\item Bits 7-8 are both 0s (optional). 
\end{packeditemize}
}
This is because the Present bit must be 1 to trigger nested page fault. Otherwise, non-present faults in the guest VM will be handled without involving the hypervisor. Bits 48-62 
are reserved and must be 0.  
The Page Size (PS) bit (bit-7) is used to determine the page size (\eg, 4KB vs. 2MB); the Global Page (G) bit (bit-8) is used to indicate whether the corresponding page is a global page.  These 2 bits can only be set 1 in the last level of the page table. Therefore, if \atknameOne generates page faults at the top-level page table, they must be set as 0.  
However, we find it possible to configure the nPT so that the first three levels of the guest page table walk all pass successfully, and only trigger the nested page fault at the last-level page table. In this way, the target memory block can be regarded as a PTE of the last-level page table and hence these two bits are not restricted to be 0s. 
It is also worth pointing out that the non-executable page-protection feature is enabled by default~\cite{amd:2019:manual}. 
For example, the level-four No-Execute (NX) bit (bit-63) controls the execution ability to execute code from all downward 128M (512 × 512 × 512 × 4KB) physical pages. The value of NX bit does not cause violation during the page table walk itself, so \atkname will succeed.

\bheading{Performance evaluation.}
The speed of memory decryption for arbitrary memory content is the same as dumping page tables, as long as the they are of PTE format. If the target block does not conform to the PTE format, a triple fault takes place instead of nested page fault, in which case the adversary could take the VMCB rewinding approach and target another memory block in the next round of attacks. 





\bheading{Percentage of readable memory blocks.} We studied the binary file of ten common applications, \texttt{python} 2.7, \texttt{OpenSSH} 7.6p1, \texttt{perl} 5.26.1, \texttt{VIM} 8.0.1453, \texttt{tcpdump} 4.9.3, \texttt{patch} 2.7.6, \texttt{grub-install} 2.02.2, \texttt{sensors} 3.4.0 , \texttt{Nginx} 1.14.0, and \texttt{diff} 3.6, which are installed from the default package archives in Ubuntu 18.04 (64-bit). The percentages of 8-byte aligned memory blocks that can be directly read using this method is 1.00\%, 1.53\%, 1.79\%, 1.81\%, 2.10\%, 3.50\%, 4.00\%, 5.88\%, 6.10\%, and 6.50\%. While they only account for a small portion of the whole memory space, they leak enough information for process fingerprinting purposes. 
\looseness=-1


\subsection{Variant 2: Executing Victim VM's Encrypted Instructions}
\label{sec:attacks:v2}





In \atknameTwo, we show that, when certain conditions are met, it is possible for the attacker VM to momentarily execute a few instructions that are encrypted in the victim VM's memory. Apparently, \atknameTwo is more powerful than the previous variant. Fortunately, the only prerequisite of \atknameTwo is the consequence of \atknameOne.



Similar to the settings in the previous attack variant, two SEV VMs were configured so that the ASID of the victim VM is 1 and the ASID of the attacker VM is 2. We assume that the attacker VM aims to execute one instruction---``{\tt movl \$2020, \%r15d}"---in the victim VM's encrypted memory. Let the virtual address of this target instruction be gVA$_0$ and the corresponding gCR3 of the target process be gCR3$_0$.
The adversary's strategy is to follow the common steps of \atkname attacks and manipulate the nPT of the attacker VM so that it finishes a few nested page table walks to successfully execute this instruction. More specifically, \atknameTwo can be performed in the following steps:

%

\bheading{\ding{192} Prepare nPT.} The hypervisor clears the Present bit of all PTEs of the attacker VM's nPT. It also prepares valid mappings for the gVA$_0$ to the physical memory encrypted with the victim's VEK. To do so, the hypervisor needs to prepare five gPA to sPA mappings (for the gPFNs of the four levels of the gPT and the instruction page), respectively.


\bheading{\ding{193} Set \nrip.} The hypervisor sets \nrip as gVA$_0$. It also clears the Interrupt Flag of the RFLAGs register (RFLAGS.IF) in the VMCB, so that the attacker VM directly executes the next instruction specified by \nrip, instead of referring to Interrupt-Descriptor-Table Register.

\bheading{\ding{194} Change ASID.} The hypervisor changes the attacker VM's ASID to the victim's ASID, marks the VMCB as dirty, and resumes the attacker VM with VMRUN. During the next VMEXIT, the value of $\%r15$ has been changed to $\$2020$, which means the attacker VM has successfully executed an instruction that is encrypted with the victim's VEK.  

These experiments suggest that \atkname allows the attacker VM to execute some instruction of the victim VM. We exploit this capability to construct decryption oracles and encryption oracles.

\subsubsection{\bf Constructing Decryption Oracles}

A decryption oracle allows the adversary to decrypt an arbitrary memory block encrypted with the victim's VEK. With \atknameTwo, the attacker VM executes one instruction of the victim VM to decrypt the target memory. 

The first step of constructing a decryption oracle is to locate an instruction in the victim VM with the format of ``{\tt mov    (\%$reg_1$),\%$reg_2$}'', which loads an 8-byte memory block whose virtual address is specified in $reg_1$ to register $reg_2$. As most memory load instructions follow this format, the availability of such an instruction is not an issue. The adversary can leverage \atknameOne to scan the physical memory of the victim VM, in hope that the readable memory blocks contain such a 3-byte instruction. Alternatively, if the kernel version of the victim VM is known, the adversary can scan the binary file of the kernel image to locate this instruction and then obtain its runtime location by reading the gPT, which can be completely extracted by \atknameOne. \looseness=-1 

Let the virtual address of this instruction be gVA$_0$, its corresponding system physical address be sPA$_0$, and the gCR3 value of the process in the victim VM be gCR3$_0$. The virtual address and the system physical address of the target memory address to be decrypted are gVA$_1$ and sPA$_1$. Note since the adversary is able to extract the gPT of the victim, the corresponding translation for gVA$_0$ and gVA$_1$ can be obtained. Then following the three steps outlined above, during a VMEXIT of the attacker VM, the adversary prepares the nPT of the attacker VM (including one mapping for gCR3$_0$, four mappings for gVA$_0$, and four mappings for gVA$_1$), configures the VMCB (including \nrip, ASID, the value of $\%reg_1$), and then resumes the attacker VM. 

In the next VMEXIT, the adversary is able to extract the secret stored in sPA$_1$ by checking the value of $\%reg_2$. The adversary can immediately perform the next round of memory decryption. The system physical page frame number can be manipulated in the last-level nPT and the page offset can be controlled in $\%reg_1$.

\bheading{Performance evaluation.}
We measured the performance of the decryption oracle described above for decrypting a 4KB memory page. With only one attacker VM, the average decryption time (of 5 trials) for a 4KB page was 113.6ms with one standard deviation of 4.3ms. 
Note the decryption speed is slower than the optimized version of \atknameOne, but the decryption oracle constructed with \atknameTwo is more powerful as it is not limited by the format of the target memory block.
\looseness=-1

\subsubsection{\bf Constructing Encryption Oracles}

An encryption oracle allows the adversary to alter the content of an arbitrary memory block encrypted with the victim’s VEK to the value specified by the adversary. With \atknameTwo, an encryption oracle can be created in ways similar to the decryption oracle. The primary difference is that the target instruction is of the format ``{\tt mov \%$reg_1$,(\%$reg_2$)}'', which moves an 8-byte value stored in $reg_1$ to the memory location specified by $reg_2$. With an encryption oracle, the adversary could breach the integrity of the victim VM and force the victim VM to (1) execute arbitrary instruction, or (2) alter sensitive data, or (3) change control flows. Note that our encryption oracle differs from those in the prior works~\cite{du:2017:sevUnsecure,buhren:2017:fault,Li:2019:sevio} as it does not rely on SEV's memory integrity flaws.



\bheading{Performance evaluation.}
We measured the performance of the encryption oracle by the time it takes to updates the content of a 4KB memory page. 
The average time of 5 trials was 104.8ms with one standard deviation of 6.1ms. 
Note in a real-world attack, the attacker may only need to change a few bytes to compromise the victim VM, which means the attack can be done within 1ms. 

\subsubsection{\bf Locating Decryption/Encryption Instructions}
In the previous experiments, we have already shown that once the instructions to perform decryption and encryption can be located, the construction of decryption and encryption oracles is effective and efficient. Next, we show how to locate such decryption/encryption instructions to bridge the gap towards an end-to-end attack. We assume the adversary has some knowledge of binary installed inside the guest VM (\eg, sshd) and its memory layout (\eg, via debugging on her own machine). 

Specifically, on the victim VM, an OpenSSH server (SSH-2.0-OpenSSH-7.6p1 Ubuntu-4ubuntu0.1) is pre-installed. 
\textit{First}, the adversary learns the version of the OpenSSH binary by monitoring the SSH handshake protocol. More specifically, the adversary who controls the hypervisor and host OS monitors the incoming network packets to the victim VM to identify the SSH \texttt{client\_hello} message. 
The victim VM would immediately respond with an SSH \texttt{server\_hello} message, which contains the version information of the OpenSSH server. As these messages are not encrypted, the adversary could leverage this information to search encryption/decryption instructions offline from a local copy of the binary.

\textit{Second}, the adversary extracts the gCR3 of the \textit{sshd} process. To do so, upon observing the \texttt{server\_hello} message, the adversary immediately clears the Present bits of all PTEs of the victim VM. The next memory access from the \textit{sshd} process will trigger an NPF VMEXIT, which reveals the value of gCR3. We empirically validated that this approach allows the adversary to correctly capture \textit{sshd}'s gCR3, by repeating the above steps 50 times and observing correct gCR3 extraction every time.


\textit{Third}, the adversary uses \atknameOne to dump a portion of the page tables of \textit{sshd} process. More specifically, the adversary first dumps the 4KB top-level page-table page pointed to by gCR3; she identifies the smallest offset of this page that represents a valid PTE, and then follow this PTE to dump the second-level page-table page. The adversary repeats this step to dump all four levels of page tables for the lowest range of the virtual address. In this way, the adversary could obtain the physical address corresponding to the base virtual address of the OpenSSH binary. 

\textit{Fourth}, with the knowledge of the memory layout of the code section of the OpenSSH binary, the adversary can calculate the physical address of the decryption/encryption instructions within the OpenSSH binary. In our demonstrated attack, the adversary targets two instructions inside the \texttt{error} function of OpenSSH, ``{\tt mov (\%rbx),\%rax}'' for decryption and ``{\tt mov \%rax,(\%r12)}'' for encryption. The offsets of the two instructions are {\tt 0xca9a} and {\tt 0xca18}, respectively. 


\bheading{Performance evaluation.}
We measured the time needed to locate these two instructions. Once the adversary has intercepted the SSH handshake messages, it takes on average 504.74ms (over 5 trials) to locate these two instructions. After locating there two instructions, the overall time to decrypt/encrypt a 4KB memory page is 504.74ms (to locate the two instructions) plus 113.6ms/104.8ms (to repeatedly execute the target instruction for decrypting/encrypting a 4KB memory page).


\subsection{Discussion on Stealthiness and Robustness}
\atkname attacks are stealthy. The attacker VM and the victim VM are two separate VMs. They have different nPTs and VMCBs and they run on different CPUs. Therefore, any execution state changes made in the attacker VM are not synchronized with the victim VM, which means 
it is impossible for victim VM to sense the presence of the attacker VM. In contrast to all known attacks to SEV, \atkname cannot be detected by running a detector in the victim VM. More interestingly, the adversary can rewind the attacker VM's VMCB to eliminate the side effects caused by the attacker VM's attack behaviors (\eg, triggering a NPF with non-PTE format or executing an illegal instruction). This method also increases the robustness of the attack: Even if the instructions of the decryption oracle are not correctly located, \atknameTwo will not affect the execution of the victim VM. Therefore, the adversary can perform the attack multiple times until it succeeds. 

\section{Applicability to SEV-ES}
\label{sec:countermeasure}

%

\subsection{Overview of SEV-ES}
\label{sec:counter:seves}

To protect VMCB during VMEXIT, SEV-ES was later introduced by AMD~\cite{kaplan:2017:seves}. With SEV-ES, a portion of the VMCB is encrypted with authentication.
Therefore, the hypervisor can no longer read or modify arbitrary register values during VMEXITs. To exchange data between the guest VM and the hypervisor, a new structure called Guest Hypervisor Control Block (GHCB) is shared between the two. The guest VM is allowed to indicate what information to be shared through GHCB. \looseness=-1

VMEXITs under SEV-ES modes are categorized into Automatic Exits (AE) and Non-Automatic Exits (NAE). AE VMEXITs (\eg, those triggered by most nested page faults, by the {\tt PAUSE} instruction, or by physical and virtual interrupts) are VMEXITs, which do not need to expose register values to the hypervisor. Therefore, AE VMEXITs directly trigger a VMEXIT to trap into the hypervisor. To enhance security, NAEs (\eg, those triggered by {\tt CPUID}, {\tt RDTSC}, {\tt MSR\_PROT} instructions) are first emulated by the guest VM instead of the hypervisor. Specifically, NAEs first trigger a \#VC exception, which is handled by the guest OS to determine which register values need to be copied into the GHCB. This NAE 
VMEXIT 
will then be handled by the hypervisor that extracts the register values from the GHCB. After the hypervisor resumes the guest in VMRUN, the \#VC handler inside the guest OS reads the results from the GHCB and copies the relevant register states to corresponding registers.

%

SEV-ES VMs can run concurrently with SEV VMs and non-SEV VMs. 
After VMEXIT, the hardware recognizes an SEV-ES VM by the SEV control bits (bit 1 and 2 of {\tt 090}h) in the VMCB~\cite{amd:2019:manual}. Therefore, the hypervisor may change the SEV type (from an SEV VM to an SEV-ES VM) during VMEXIT. The legal ASID ranges of SEV-ES and SEV VMs, however, are disjoint, and thus it is not possible to run an SEV-ES VM with an ASID in the range of SEV VMs.

\bheading{VMCB's Integrity Protection.} With SEV-ES, the original VMCB is divided into two separate sections, namely the control area and the state save area (VMSA)~\cite{amd:2019:manual}. The control area of VMCB is unencrypted and controlled by the hypervisor, which contains the bits to be intercepted by the hypervisor, the guest ASID ($058$h), control bits of SEV and SEV-ES ($090$h), TLB control ($058$h), VMCB clean bits ({\tt 0C0}h), \nrip ({\tt 0C8}h), the gPA of GHCB ({\tt 0A0}h), the nCR3 ({\tt 0B0}h),  VMCB save state pointer ({\tt 108}h), \etc The state save area is encrypted and integrity protected, which contains the saved register values of the guest VM. 
The VMCB save state pointer stores the system physical address of VMSA---the encrypted memory page storing the state save area. 

The integrity-check value of the state save area is stored in the protected DRAM, which cannot be accessed by any software, including the hypervisor~\cite{amd:2019:manual}. At VMRUN, the processor performs an integrity check of the VMSA. If the integrity check fails, VMRUN terminates with errors~\cite{amd:2019:manual}. Because the integrity-check value (or the physical address storing the value) is not specified by the hypervisor at VMRUN, we conjecture the value is index by the system physical address of the VMSA. Therefore, a parked virtual CPU is uniquely identified by the VMSA physical address.




\subsection{\atknameOne on SEV-ES}
There are two main challenges when applying \atkname to SEV-ES.
The primary challenge is to bypass the VMSA check. Directly resuming the attacker VM using the victim's ASID would cause VMRUN to fail 
immediately, because the VMSA integrity check takes place before fetching any instructions in the attacker VM. Since the attacker VM's VMSA is encrypted using the VEK of the attacker VM, when resuming the attacker VM with the victim's ASID, the decryption of VMSA leads to garbage data, crashing the attacker VM immediately. 
Therefore, to perform \atknameOne, the adversary must change the save state pointer (0108h) of the attacker VM's VMCB so that the attacker VM will reuse the victim VM's VMSA. 

The second challenge is to control the decrypted memory block's page offset. As the attacker VM reuses Victim VM's VMSA, the attacker VM cannot change the register values that are stored in the VMSA, which includes \rip, gCR3, and all general-purpose registers (if not exposed in the GHCB). Therefore, with SEV-ES, the adversary is no longer able to directly control the execution of the attacker VM by simply manipulating its \nrip in its VMCB's control area~\cite{amd:2019:manual}. 
However, 
by pausing victim's VCPU at different execution points,  
the \nrip can be different at each VMEXIT. As such, the adversary is still able to perform \atknameOne on SEV-ES VMs to achieve the same goal---extracting the entire gPT or decrypting any 8-byte memory block conforming to a PTE format. To show this, we have performed the following experiments: \looseness=-1

Two SEV-ES VMs were launched. The ASID of the victim VM is set to be 1 and that of the attacker VM is 2. The hypervisor pauses the victim VM at one of its VMEXITs, so that its VMSA is not used by itself. The attack is performed in the following steps: \looseness=-1

\bheading{\ding{192} Prepare nPT.} During the VMEXIT of the attacker VM, the hypervisor clears all the Present bits in the nPT of the attacker VM. 


\bheading{\ding{193} Manipulate the attacker VM's VMCB.} The hypervisor first changes the attacker VM's ASID from 2 to 1. It also informs the hardware to flush all TLB entries of the current CPU, by setting the TLB clearing field (058h) in the VMCB control area.
Finally, it changes the VMCB save area pointer to point to the victim's VMSA.

\bheading{\ding{194} Resume the attacker VM.} Because the attacker VM runs with the victim's ASID, the victim's VMSA is decrypted correctly. The integrity check also passes, as no change is made in the VMSA, including its system physical address. Once resumed, the attacker VM will try to fetch the first instruction determined by \rip (in VMSA) or the IDTR using the victim's VEK. Since there is no valid TLB entry, the processor has to perform a guest page table walk to translate the virtual address to the system physical address. A nested page fault can be observed with the faulting address being the victim VM's gCR3 value.

\bheading{\ding{195} Remap gCR3 in nPT.} When handling this NPF VMEXIT, the hypervisor remaps the gCR3 in the nPT to the victim VM's memory page to be decrypted. The Present bits of the corresponding nested PTEs are set to avoid another NPF of this translation. Moreover, the {\tt EXITINTINFO} field in the unencrypted VMCB control area needs to be cleared to make sure the attacker VM complete the page table walk. After resuming the attacker VM, an NPF for the translation of another gPA (embedded in the target memory block) will occur, which reveals the content of the 8-byte aligned memory block if it 
conforms to the PTE format. 


\bheading{\ding{196} Reuse the VMSA.} The hypervisor repeats step \ding{195} so that its gCR3 is remapped to the next page to be decrypted in the victim VM. Then, the next NPF VMEXIT reveals the corresponding memory block. This could work because the attacker VM has not successfully fetched a single instruction yet; it is trapped in the first page table walk (more specifically, the top-level 
nested page table walk of the first gPA). Therefore, the VMSA is not updated and no valid TLB entry is created. During the remapping of gCR3, the hypervisor is able to invalidate the previously generated entry in the nTLB. Thus, from the perspective of the attacker VM, step \ding{195} does not change its state. Therefore, the attacks can be carried out repeatedly. \looseness=-1


\bheading{\ding{197} Handling triple faults.} In step \ding{195} or step \ding{196}, if the targeted 8-byte memory block does not conform to the PTE format, a triple fault VMEXIT (error code 0x7f) will be triggered instead of the NPF VMEXIT. The adversary can continue to decrypt the next page if this happens. However, after a triple fault, the \rip in the VMSA has been updated to the fault handler to deal with the fault. As such, resuming from a triple fault will lead to the decryption of a different offset of the target page. Nevertheless, the attack can still continue.

\subsubsection{Resuming the Victim VM} 
After performing \atknameOne, the VMSA of the victim VM is still usable by the victim. We empirically validated this by resuming the victim VM after the attacker VM has used this VMSA to decrypt several memory blocks and has encountered both nested page faults and triple faults. The victim VM was resumed successfully, without observing any faults or abnormal kernel logs (as discussed in \secref{sec:es-stealthiness}).
To better understand the victim VM's state changes when its VMSA is used by the attacker VM, we 
checked which regions of the encrypted VMSA's ciphertext blocks have been changed after the attacker VM has performed several rounds of \atknameOne, which triggers both nested page faults and triple faults. The result shows that the entire VMSA remains the same, except the value of CR2, which stores the most recent faulting address. The change of the CR2 value does not affect the execution of the victim VM as this value is not used by the guest OS after NPFs. 

\subsubsection{Controlling Page Offsets}
Because the integrity protection of VMSA prevents the adversary from controlling the \rip after VMRUN, the page offset of the memory blocks to be decrypted cannot be controlled on SEV-ES. However, the adversary may resume the victim VM and allow it to run till a different \rip is encountered. In total, 512 different \rip{s} are needed to decrypt any memory blocks conforming to the PTE formats. 
Two challenges remain: First, under an unknown \rip, how can the hypervisor determine the page offset of the memory blocks to be decrypted; second, how to diversify the \rip{s} in order to cover more offsets.

\begin{figure}[t]
\centering
\begin{subfigure}[b]{0.4\columnwidth}
\includegraphics[width=\textwidth]{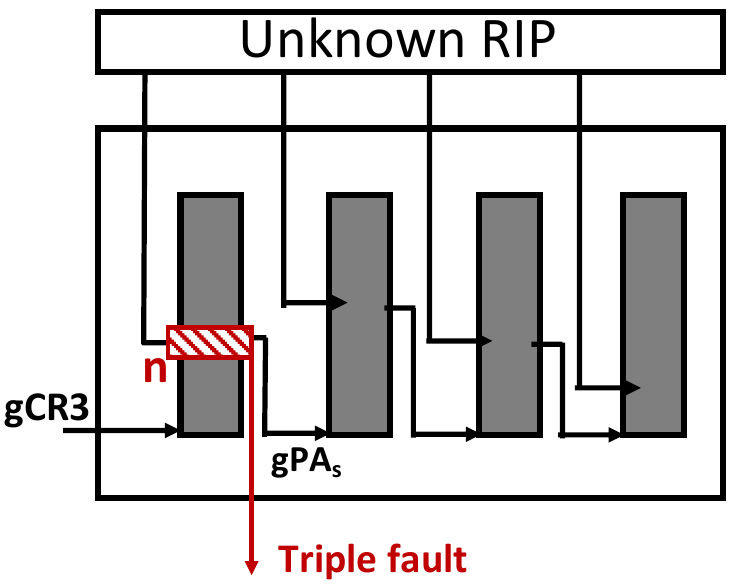}
\caption{Determine RIP's offset.}
\label{fig:offset_flow}
\end{subfigure}
\hfill
\begin{subfigure}[b]{0.59\columnwidth}
\includegraphics[width=\textwidth]{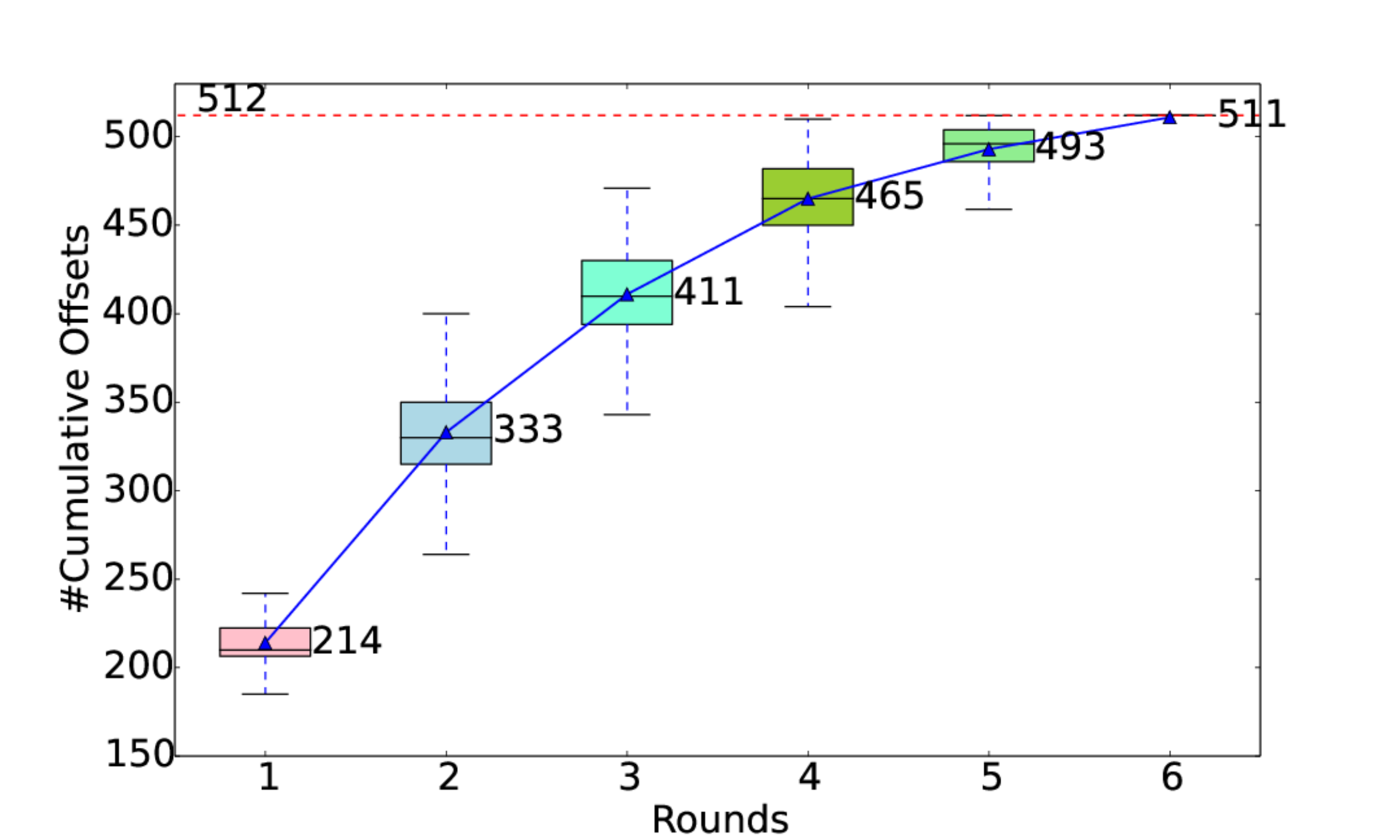}
\caption{Covered offsets after $N$ rounds.}
\label{fig:c6_offsets}
\end{subfigure}
\caption{Controlling page offsets.}
\label{fig:controll_offsets}
\end{figure}

First, to determine the corresponding page offset for an unknown \rip, the hypervisor may adopt the following approach, as shown in \figref{fig:offset_flow}. In the first step, the adversary obtains the physical address of one of the victim VM's last-level page table page. This can be achieved by clearing the Present bits of all pages and observing the subsequent NPFs: The faulting address of the first NPF reveals the value of gCR3 of the current process inside the victim VM and the faulting address of the fourth NPF reveals the address of a last-level page table page. It is preferred that this last-level page table page is not actively used by the victim; otherwise fault may occur inside the victim VM. In the second step, the hypervisor remaps the victim VM's gCR3 value obtained in step \ding{194} to this last-level page table page, and then performs \atkname V1 to extract the value of the PTE entry corresponding to the current RIP. Let us assume the offset of this PTE entry is $n$ and extracted value is gPA$_s$. In the third step, the adversary directly modifies the ciphertext of this last-level page table and perform \atkname V1 again. If the change includes offset $n$, \atkname will likely encounter a triple fault as the target block does not conform to the PTE format after decryption, or in some cases extract a value that is different from gPA$_s$. Otherwise, \atkname will extract the same value gPA$_s$. Using this primitive, the adversary can perform either a binary search or a simple linear search on the targeted page table page, eventually revealing the value of the offset $n$. In our experiments with over 200 trials, it takes 19.28ms on average to determine the offset of a \rip. Note that to avoid crashing the victim processes, the adversary should change the ciphertext of the page table page back to the original value.

To diversify the exploited \rip{s}, one strategy is to pause the victim when the VMEXIT is a NPF-triggered AE. When VMEXITs are NAEs or interrupt-triggered AEs, the next instruction to be executed after VMRUN is an instruction of the \#VC handler, whose virtual address is fixed in the kernel address space. To differentiate NPF-triggered AEs and interrupt-triggered AEs, although the adversary cannot read the RFLAG.IF directly, which indicates pending interrupts, she can inspect Bit 8 (V\_IRQ) of the Virtual Interrupt Control field (offset 60h) in the unencrypted VMCB control area. Moreover, as two consecutive NPF-triggered AEs may be caused by the same \rip, it is preferred to pause the victim VM after a few AEs. To trigger more NPF VMEXITs, one could periodically unset the Present bit of all PTEs of the victim VM. 

With these strategies in place, we empirically evaluated the time needed for the adversary to find all 512 offsets. In our test, we let the victim VM run a build-in program of Ubuntu Linux, called ``cryptsetup benchmark". 
The attack can be performed on any level of the page tables; bits 47-39, 38-30, 29-21, and 20-12 of the same RIP can all be used as the page offset by the attacker. Therefore, with any RIP, there are 1$\sim$4 different offsets that the attacker may use to extract data on any encrypted page. 
The experiments were performed in the following manner: Each round of the experiments, the cryptsetup benchmark were run several times and each time with a different address space layout due to ASLR;  every 30 seconds, the adversary unset all Present bits of the victim VM to trigger NPFs; the adversary pauses the victim VM every 13 AE VMEXITs to extract one \rip. The adversary concludes the round of monitoring after 60 seconds. In total, 15 rounds of experiments were conducted. 
\figref{fig:c6_offsets} shows the number of offsets that can be covered after N rounds of experiments, where N=1 to 6. Each data point is calculated over all combinations of selecting N rounds from the 15 rounds, \ie,  
C(15, N), of data collected in the experiments above.
Specifically, on average, after 5 rounds of experiments, the adversary could obtain 493 offsets; after 6 rounds, she could obtain 511 offsets (out of the 512 offsets). These experiments show that when the victims run an application that has diverse \rip{s} (\ie, not running in idle loops), the adversary has a good chance of performing \atknameOne on almost all page offsets after some efforts (in these experiments, after 6 minutes of the victim's execution).

\subsubsection{Performance Evaluation}
%
We have evaluated the attack mentioned above on a workstation with an 8-Core AMD EPYC 7251 Processor. The motherboard of our testbed machine was GIGABYTE MZ31-AR0, with which we successfully configured Fn8000\_001F[EDX] to return 5, which means ASID 1 to 4 were reserved for SEV-ES VMs.  Since the source code supporting SEV-ES for both host OS and guest OS has not been added into the mainstream Linux kernel yet, we used the source code provided in the SEV-ES branch of AMD's official repositories for SEV, which is available on Github~\cite{amd:2020:github}. The kernel version for the host and guest were branch sev-es-5.1-v9. The QEMU version used was QEMU sev-es-v4 and the OVMF version was sev-es-v11.
Both victim VMs and attacker VMs were configured as SEV-ES-enabled VMs with 1 virtual CPU, 2 GB DRAM and 30 GB disk storage. All VMs were created by the kernel image generated from sev-es-5.1-v9 branch without any additional modification.


On average over 200 trials, it takes 2.0ms to decrypt one 8-byte memory block, which is slower than the attack against SEV VMs (0.077ms per block). This is because the AMD-SP must calculate the hash of the VMSA and store it to the secure memory region during VMEXITs, and validate its integrity after each VMRUN. This happens in between of decrypting two memory blocks.

\subsection{Discussion on Stealthiness}
\label{sec:es-stealthiness}
To attack SEV-ES VMs, the attacker VM must reuse the victim VM's VMSA. However, \atknameOne is still stealthy and undetectable by the victim VM for three reasons. First, the attack only alters the CR2 field of the victim's VMSA. As this field is not examined by the guest OS after resumption from a NPF, the victim VM cannot detect the anomaly. Second, even if the guest OS is modified to monitor CR2, the change of CR2 cannot be detected, because the AE NPFs are directly trapped into the hypervisor, such that the guest OS does not have a chance to record the original value of CR2 to be compared with. Third, the attacker can perform the following steps to confuse the detector: Every time an \atkname attack is performed, the attacker could “clean up” the trace by forcing a NPF on the victim’s next instruction. In this way, even if the victim can observe CR2 changes, CR2 is filled with a “normal” page faults. The victim will not observe unexpected “abnormal” CR2 values.

\subsection{\atknameTwo on SEV-ES}

Applying \atknameTwo on SEV-ES would be challenging, because with the encrypted VMSA, \rip is no longer controlled by the adversary. As such, the attacker VM will resume from the \rip stored in the VMSA, which prevents the attacker VM from executing arbitrary instructions. Moreover, constructing useful encryption or decryption oracles requires the manipulation of specific register values, which is only possible without SEV-ES. 
\looseness=-1

\ignore{
\begin{figure}[t]
\centering
\includegraphics[width=0.9\columnwidth]{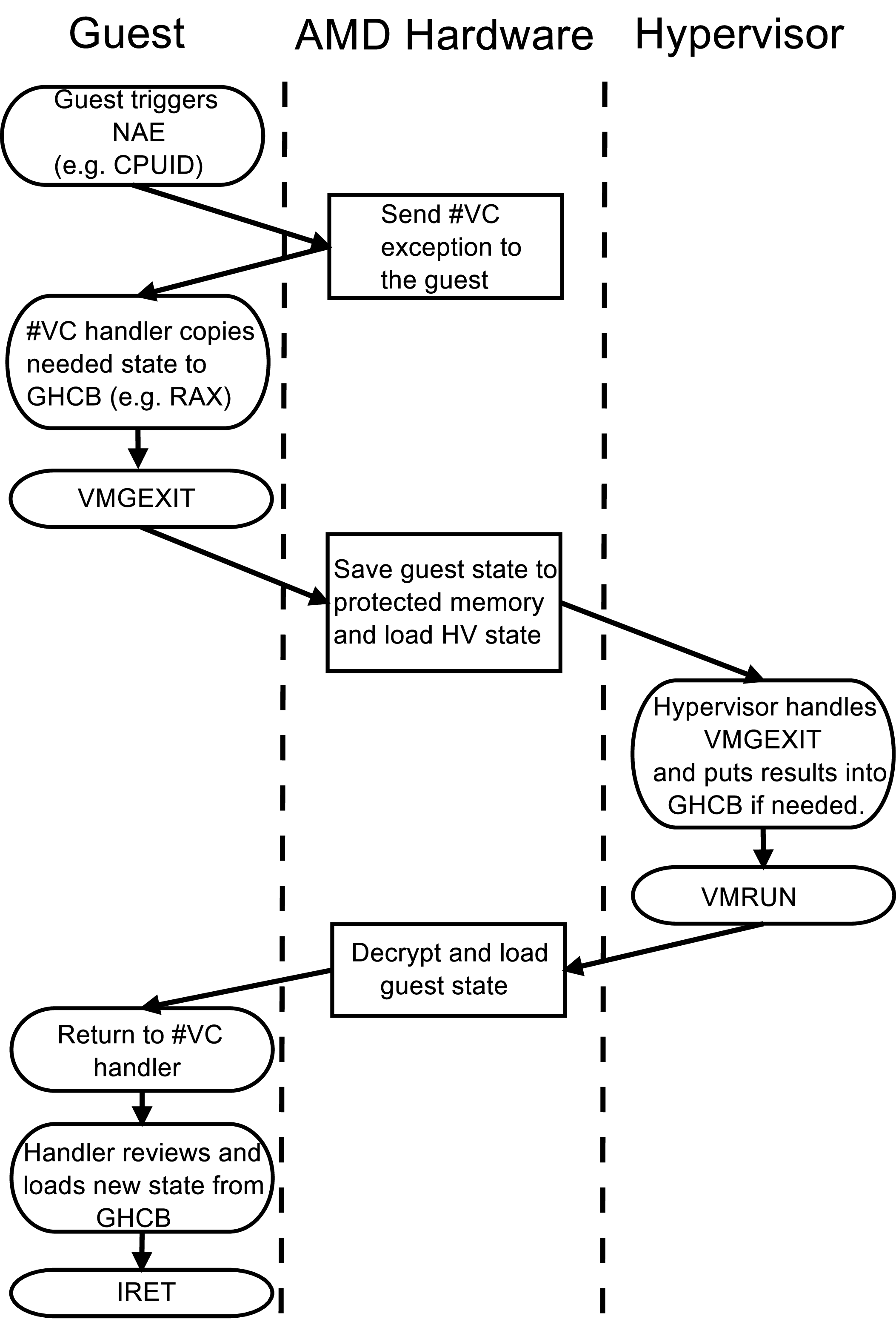}
\caption{NAE example flow .}
\label{fig:NAE}
\end{figure}
}







\section{Discussion}
\label{sec:discuss}


\subsection{Applicability to SEV-SNP}

To address the attacks against SEV that exploit memory integrity flaws, AMD recently announced SEV-SNP~\cite{david:2019:sevsnp} and released a whitepaper describing its high-level functionality in January, 2020~\cite{amd:2020:snp}. 
The key idea of SEV-SNP is to provide memory integrity protection using a Reverse Map Table (RMP). An RMP is a table indexed by system page frame numbers. One RMP is maintained for the entire system. Each system page frame has one entry in the RMP, which stores information of the page state (\eg, hypervisor, guest-invalid, guest-valid) and ownership (\ie, the VM's ASID and the corresponding gPA) of the physical page. The ownership of a physical page is established through a new instruction, {\tt PVALIDATE}, which can only be executed by the guest VM. Therefore, the guest VM can guarantee that each guest physical page is only mapped to one system physical page; by construction, RMP allows each system physical page to have only one validated owner. \looseness=-1

After each nested page table walks that leads to a system physical page belonging to an SEV-SNP VM (and also some other cases), an RMP check is to be performed. The RMP check compares the owner of the page (\ie, the ASID) with the current ASID and compares the recorded gPA in the RMP entry with the gPA of the current nPT walk. If a mismatch is detected, a nested page fault will be triggered. 

\begin{packeditemize}

\item \textit{\bf \atknameOne on SEV-SNP.}
When applying \atknameOne on SEV-SNP by following the same attack steps for SEV-ES, it seems step \ding{192} to \ding{195} would work the same. As the VMSA is also protected by the RMP, loading VMSA would lead to an RMP check. 
However, as the attacker VM uses the victim's ASID, the check would pass. However, the NPF in step \ding{196} that reveals the page content would not occur. Instead, an NPF due to RMP check would take place, because the gPA used in nPT walk is different from the one stored in the RMP entry. Therefore, from the description of the RMP, it seems \atknameOne can be prevented. 

\item \textit{\bf \atknameTwo on SEV-SNP.}
As \atknameTwo does not work on SEV-ES, it cannot be applied on SEV-SNP. \looseness=-1

\end{packeditemize}



\subsection{Real-world Impact}
\atkname can be more damaging to the SEV-based cloud industry than other known attacks. For instance, Google Cloud recently provides 
SEV-enabled VMs, called Confidential VMs, as its first product of Confidential Computing~\cite{google:2020:sev}. \atkname attacks are the only attacks that are undetectable by the victim VM. Therefore, it is possible for a malicious insider to peak into the encrypted memory without being noticed by Google or the cloud user.  

\subsection{Relation to Speculative Execution Attacks}

\atkname is \textit{not} a speculative execution attack. Meltdown~\cite{lipp:2018:meltdown}, Spectre~\cite{kocher:2019:spectre}, L1TF~\cite{van:2018:foreshadow}, and MDS~\cite{van:2019:ridl,schwarz:2019:zombieload,canella:2019:fallout} are prominent speculative execution attacks that exploit transiently executed instructions to extract secret memory data through side channels. In these attacks, instructions are speculatively executed while the processor awaits resolution of branch targets, detection of exceptions, disambiguation of load/store addresses, \etc. However, in the settings of \atknameOne, no instructions are executed, as the exceptions take place as soon as the frontend starts to fetch instructions from the memory. \atkname V2 executes instructions with architecture-visible effects.

\atkname does not rely on micro-architectural side channels, either. Speculative execution attacks leverage micro-architectural side channels (\eg, cache side channels) to leak secret information to the program controlled by the attacker. In contrast, \atkname reveals data from the victim VM as page frame numbers, which can be learned by the hypervisor directly during page fault handling.

\subsection{Yet Another \atkname Variant: Reusing Victim's TLB Entries}
\label{sec:discuss:v3}

We next present another variant of \atkname, which allows the attacker VM to reuse the TLB entries of the victim VM for address translation and execute some instructions, even without any successful page table walks.

Two VMs are involved in a proof-of-concept attack: the victim VM is an SEV VM whose ASID is 1; the attacker VM is a non-SEV VM whose ASID is 16. Both VMs only have one VCPU, which are configured by the hypervisor to run on the same logical CPU core. We assume the victim VM executes the following instructions: 

\begin{lstlisting}[language=Python,basicstyle=\small]
 d83:41 bb e4 07 00 00   mov    $0x7e4,%r11d
 d89:41 bc e4 07 00 00   mov    $0x7e4,%r12d
 d8f:0f a2               cpuid  
 d91:eb f0               jmp    d83 
\end{lstlisting}

Specifically, the code updates the values of {\tt \%r11d} and {\tt \%r12d}, and then executes a {\tt CPUID} to trigger a VMEXIT. Following the common steps of \atkname, the adversary launches an attacker VM, changes its ASID during VMEXIT, sets the \nrip of the attacker VM to the virtual address of the code snippet above, changes offset 090h of VMCB to make it an SEV VM, and resumes the attacker VM. Unlike \atknameOne and \atknameTwo, the nPT of the attacker VM is not changed in this step. Therefore, if the attacker VM performs a page table walk, a NPF will be triggered.\looseness=-1

Interestingly, the execution of the attacker VM triggers {\tt CPUID} VMEXITs before a triple fault VMEXIT crashes it. Since no NPF is observed, the attacker VM apparently does not perform any page table walk.
However, during the attacker VM's  {\tt CPUID} VMEXITs, we observe that the values of {\tt \%r11d} and {\tt \%r12d} have been successfully changed to {\tt \$0x7e4}. It is clear that the two MOV instructions and the subsequent  {\tt CPUID} instruction have been executed by the attacker VM. This is because the attacker VM was able to reuse the victim VM's TLB entries to translate the virtual address of the instructions. 

While the consequences of this attack are close to V2, it highlights the following flaws in AMD's TLB isolation between guest VMs: (1) ASIDs serve as the only identifier for access controls to TLBs, which can be \textit{forged} by the hypervisor, and (2) TLBs cleansing during VM context switch is performed at the discretion of the hypervisor, which may be \textit{skipped} intentionally. Nevertheless, it is fair to note constructing a practical end-to-end attack using this attack variant is still difficult to accomplish.

\section{Related Work}
\label{sec:related}
\begin{table*}[t]
\centering
\caption{\small Demonstrated attacks against SEV. I/O Interaction: the attack requires interaction with applications inside the victim VM through I/O operations (\eg, Network, disk). 
Stealthiness: the attack cannot be detected by the victim VM. 
}
{\footnotesize
\begin{tabular}[t]{c?c?c?c?c?c?c}
\Xhline{1pt}
\textbf{Research Papers}  & \textbf{Exploited Vulnerabilities} & \textbf{I/O Interaction}  &\textbf{\tabincell{c}{Breach \\ Confidentiality}} &\textbf{\tabincell{c}{Breach \\ Integrity}} &\textbf{Stealthiness} &\textbf{Mitigated by} \\
\Xhline{1pt}
Du \etal~\cite{du:2017:sevUnsecure} & Unauthenticated encryption &\checkmark &\ding{55} & \checkmark  &\ding{55} & SEV-SNP\\ \Xhline{0.2pt}
Buhren \etal~\cite{buhren:2017:fault}  & Unauthenticated encryption & \checkmark &\checkmark &\ding{55} &  \ding{55}& SEV-SNP \\\Xhline{0.2pt}
Wilke \etal~\cite{wilke:2020:sevurity}& Unauthenticated encryption & \checkmark &\checkmark &\checkmark &  \ding{55}& SEV-SNP \\\Xhline{0.2pt}
Werner \etal~\cite{werner:2019:severest} & Unencrypted VMCB &\checkmark &\checkmark &\ding{55} &  \ding{55}& SEV-ES \\ \Xhline{0.2pt}
Hetzelt \& Buhren~\cite{hetzelt:2017:security} & \tabincell{c}{Unencrypted VMCB\\Unprotected PT} &\checkmark&\checkmark &\checkmark & \ding{55} & SEV-SNP \\ \Xhline{0.2pt}
Morbitzer \etal~\cite{morbitzer:2018:severed} & Unprotected PT &\checkmark &\checkmark &\ding{55} &\ding{55}& SEV-SNP \\\Xhline{0.2pt}
Morbitzer \etal~\cite{Morbitzer:2019:extract} & Unprotected PT & \checkmark &\checkmark &\ding{55} &\ding{55}& SEV-SNP \\\Xhline{0.2pt}
Li \etal~\cite{Li:2019:sevio} & \tabincell{c}{Unprotected I/O \\Unauthenticated encryption}&  \checkmark&\checkmark &\checkmark &\ding{55}& SEV-SNP \\\Xhline{0.2pt}
Li \etal~\cite{li:2021:cipherleaks} & \tabincell{c}{Ciphertext accessibility}&  \checkmark&\checkmark &\ding{55} &\checkmark & Hardware Patch \\\Xhline{0.2pt}
\atknameOne & Security-by-Crash & \ding{55}&\checkmark &\ding{55} &\checkmark& SEV-SNP \\ \Xhline{0.2pt}
\atknameTwo & \tabincell{c}{Security-by-Crash \\Unencrypted VMCB}&\ding{55}&\checkmark &\checkmark &\checkmark& SEV-ES\\ \Xhline{1pt}
\end{tabular} 
}
\begin{minipage}{0.59\columnwidth}
\centering
\end{minipage} 
\vspace{5pt}
\label{table:attacks}
\end{table*} 

Past work mainly studied the insecurity of AMD SEV from the following aspects.  

\bheading{Unencrypted VMCB.}
Before SEV-ES, VMCB is not encrypted during VMEXIT. Hetzelt and Buhren~\cite{hetzelt:2017:security} first reported that an adversary who controls the hypervisor could directly observe the machine states of the guest VM by reading the VMCB structure. Moreover, they show that the adversary could also manipulate the register values in the VMCB before resuming the guest VM to perform return-oriented programming (ROP) attacks~\cite{Shacham:2007:GIF} against the guest VM. As a result, the adversary is able to read or write arbitrary memory in the SEV VM. These security issues have been completely mitigated by SEV-ES~\cite{kaplan:2017:seves}. Werner \etal also explored security vulnerabilities caused by unencrypted VMCB~\cite{werner:2019:severest}. Their study suggests that an adversary is able to identify applications running inside the SEV VMs by recording register values in VMCB. The study also shows that it is practical to inject data by locating certain system calls and modify some registers to mislead the guest VM. However, SEV-ES restricts most of their attacks and the only working attack that remains is application fingerprinting.

\bheading{Unauthenticated encryption.}
The lack of authentication in the memory encryption is one major drawback of the SME design, which has been demonstrated in fault injection attacks~\cite{buhren:2017:fault}. SEV inherits this security issue. Therefore, a malicious hypervisor may alter the ciphertext of the encrypted memory without triggering faults in the guest VM. Another problem with SME's memory encryption design is that SME uses Electronic Codebook (ECB) mode of operation with an additonal tweak function in its AES-based memory encryption.
This design choice unfortunately has enabled chosen plaintext attacks. Du \etal~\cite{du:2017:sevUnsecure} reverse-engineered the tweak function and recovered the mapping between the system physical address and the output of the tweak functions.
Wilke \etal~\cite{wilke:2020:sevurity} further studied the Xor-Encrypt-Xor (XEX) mode of memory encryption of AMD's Epyc 3xx1 series processors, where the tweak function XOR with the plaintext twice, both before and after the encryption. However, 
the entropy of the tweak functions is only 32 bits, making brute-force attacks practical. It is demonstrated that the adversary who breaks the tweak function can insert some arbitrary 2-byte instruction into encrypted memory with the help of 8MB plaintext-ciphertext pairs. 
Fortunately, the XEX tweak function vulnerability exploited in the paper was fixed after Zen 2 architecture that was released in May, 2019.


 

\bheading{Unprotected nPT.}
Hetzelt and Buhren~\cite{hetzelt:2017:security} demonstrated address translation redirection attacks (an idea first explored by Jang \etal in the context of hardware-based external monitors~\cite{jang:2014:atra}) in SEV and discussed remapping guest pages in the nPT to replay previously captured memory pages. This idea was later realized by SEVered~\cite{morbitzer:2018:severed, Morbitzer:2019:extract}, which manipulates the nPT to breach the confidentiality of the memory encryption. More specifically, in the SEVered attack, the hypervisor triggers activities of the victim VM's network-facing application and concurrently monitor its accesses to the encrypted memory using a page-level side channel. In this way, the hypervisor can determine the system physical page used to store the response data. Then, by changing the memory mapping in the nPT, the hypervisor tricks the guest VM to respond to network requests from the target page, leaking secrets to the adversary.


\bheading{Unprotected I/O.}
Li \etal~\cite{Li:2019:sevio} exploited unprotected I/O operations to construct encryption and decryption oracles that encrypts and decrypts arbitrary memory with the victim's VEK. As SEV's IOMMU hardware can only support DMA with hypervisor's VEK, a shared region within SEV VM called Software I/O Translation Lookaside Buffer (SWIOTLB) is always needed for SEV I/O operations. SEV VM itself needs to copy I/O streaming from SWIOTLB to its private memory when there are incoming I/O data; it needs to copy I/O data to the SWIOTLB when there are outgoing I/O.  This design gives the hypervisor an opportunity to monitor and alternate I/O streaming to build encryption and decryption oracles. The paper also showed these unprotected I/O problems still exist in SEV-ES. 

\bheading{Ciphertext accessibility.} Li \etal~\cite{li:2021:cipherleaks} presents the first attacks against SEV-SNP. Specifically, the Cipherleaks attack is a novel side channel attack on SEV platform, in which the attackers continuously monitor the ciphertext changes in the VMSA region to infer the internal register states. The Cipherleaks attack has been applied on the state-of-the-art OpenSSL library to steal RSA private key and ECDSA nonce. Microcode patches have been released to mitigate the ciphertext side channels. 

\looseness=-5

\bheading{Summary.}
We summarize the attacks against SEV, their exploited vulnerabilities, the attack consequences, and the stealthiness of the attacks in \tabref{table:attacks}. 
SEV-SNP can defeat all known attacks against these design flaws, including unencrypted VMCB, unauthenticated encryption, unprotected nPT, and unprotected I/O. However, as SEV-SNP is not designed to mitigate ASID abuses and the \atkname attacks, while it prevents \atknameOne as it disallows nPT remapping, we plan to further investigate other forms of \atkname against SEV-SNP in the future work. \looseness=-1



\section{Conclusion}
\label{sec:conclude}

In conclusion, this paper demystifies AMD SEV's ASID-based isolation for encrypted memory pages, cache lines, and TLB entries.  For the first time, it challenges the “security-by-crash” design philosophy taken by AMD. It also proposes the \atkname attacks, a novel class of attacks against SEV that allow the adversary to launch an attacker VM and change its ASID to that of the victim VM to impersonate the victim. Two variants of \atkname attacks have been presented and successfully demonstrated on SEV machines. They are the first SEV attacks that do not rely on SEV's memory integrity flaws. 

\section*{Acknowledgement}
We thank David Kaplan and other engineers of AMD's SEV team for their valuable feedback and constructive suggestions, which have helped improve this paper. This work was in part supported by NSF Award 1750809, 1834213, and 1834216.
\looseness=-1

\bibliographystyle{ACM-Reference-Format}
\bibliography{paper}





\end{document}